\documentclass[11pt,A4paper]{JHEP3}
\usepackage{graphicx}
\usepackage{amsfonts}
\usepackage{amssymb}
\usepackage{amsmath}
\usepackage{epsfig}
\newcommand{\be}{\begin{equation}}
\newcommand{\ee}{\end{equation}}
\newcommand{\bea}{\begin{eqnarray}}
\newcommand{\eea}{\end{eqnarray}}

\title{Renormalization in General Gauge Mediation}
\author{Jae~Yong~Lee\\
School of Physics, KIAS, Seoul 130-722, Korea and\\
Department of Physics, Korea University, Seoul 136-701, Korea\\
E-mail: \email{littlehigg@kias.re.kr}}
\preprint{KIAS-P09067}
\abstract{We revisit General Gauge Mediation (GGM)
in light of the supersymmetric (linear) sigma model by utilizing the current superfield.
The current superfield in the GGM is identified with supersymmetric extension of the vector symmetry current
of the sigma model while spontaneous breakdown of supersymmetry
in the GGM corresponds to soft breakdown of the axial vector symmetry of the sigma model.
We first derive the current superfield from the supersymmetric linear sigma model
and then compute 2-point functions of the current superfield using the (anti-)commutation relations of the
messenger component fields.
After the global symmetry are weakly gauged, the 2-point functions of the current superfield
are identified with a part of the 2-point functions of the associated vector superfield. 
We renormalize them by dimensional regularization and show that physical gaugino and sfermion masses of the MSSM
are expressed in terms of the wavefunction renormalization constants of the component fields of the vector superfield.}

\begin{document}

\section{Introduction}
As the first proton-proton collisions at the LHC occurred
expectations for a new physics at a TeV scale has escalated tremendously.
One of the best candidates for the new physics at a TeV scale
is low energy supersymmetry, in particular the Minimal Supersymmetry Standard Model (MSSM).
But supersymmetry is undoubtedly broken in real world
so that the gauginos, squarks and sleptons of the MSSM get massive.
Supersymmetry must be spontaneously broken in the hidden sector,
and then its breaking is communicated to the visible sector,
yielding soft terms including the gaugino, squark and slepton masses.

There are various mechanisms for generating the soft terms. 
One of the mechanisms is Gauge Mediation (GM) where messenger chiral superfields,
charged under the SM gauge symmetry, have their scalar masses split
due to spontaneous supersymmetry breakdown in the hidden sector.
Sequently the scalar mass splitting gives a mass to the gauginos, squarks and sleptons of
the MSSM by quantum loops~\cite{Dine:1981rt,Dimopoulos:1981au,Dine:1981gu,Nappi:1982hm,
AlvarezGaume:1981wy,Dimopoulos:1982gm,
Dine:1993yw,Dine:1994vc,Dine:1995ag}.
Recently a new interpretation of the GM was introduced by Meade and his collaborators, and
it was dubbed General Gauge Mediation (GGM)~\cite{Meade:2008wd}.

There are two key steps to actualize the GGM.
The first step is to replace the messenger fields with current superfield formed by them.
As a result, the gaugino and sfermion masses are expressed in terms of the current correlators.
The second step is to regard the SM gauge symmetries as global symmetries at first
and gauge the global symmetry later.
This process is justified by the smallness of the SM gauge couplings at the relevant energy scale.
The GGM encompasses not only models with messengers but also strongly-coupled direct GM models
so that we can pinpoint features of the entire class of GM models and distinguish them
from specific signatures of various subclasses.

In this letter, we interpret an inceptive model of the GGM as a supersymmetric sigma model
with a conserved global vector symmetry.
Supersymmetric masses of messengers in the GGM do not break the vector symmetry.
Neither does messenger scalar mass splitting,
stemming from spontaneous breakdown of supersymmetry in the hidden sector,
break the vector symmetry.
Whenever a global internal symmetry is conserved there is a corresponding current associated with it.
For supersymmetric theories, the corresponding current forms a supermultiplet
which is so called the linear superfield.

We revisit the current superfield of the GGM~\cite{Meade:2008wd}.
We first derive components of the current superfield from K\"{a}hler potential
and then elucidate the relation between current superfield and supercharge~\cite{Buican:2008ws}.
We explicitly derive the (anti-)commutation relations between its components
and supercharge from quantum field theories,
stressing that the (anti-)commutation relations are a consequence
both of supersymmetry and of conservation of the global vector symmetry.
Furthermore, we stress that these relations still hold even if supersymmetry is broken spontaneously.

The authors in ref.~\cite{Meade:2008wd,Buican:2008ws} obtain a few invariant functions
from the 2-point functions, and derive universal soft masses from them.
The authers in ref.~\cite{Buican:2009vv} discuss these functions in perspective of effective field theory.
In this letter, we further develop their rationale by taking account of a renormalization procedure.
The invariant functions are nothing but the wavefunction renormalization constants of a vector superfield,
so we can specify the connection between the GGM and the wavefunction renormalization procedure
in ref.~\cite{Giudice:1997ni,ArkaniHamed:1998kj}. 

The outline of this paper is briefly listed at the top of the first page.

\section{Supersymmetric linear sigma model}\label{sect2}
\subsection{Current superfield}
This section is mostly adopted from the exposition of S. Weinberg~\cite{Weinberg:2000cr}.
We consider a Wess-Zumino model with messenger chiral superfields whose components are expressed
in terms of $x, \theta$ and $\bar\theta$:
\be
\begin{split}
\boldsymbol{\phi}_b&=\phi_b+\sqrt{2}\theta\psi_b+\theta\theta F_b+i\theta\sigma^\mu\bar\theta\partial_\mu\phi_b
-\frac{i}{\sqrt{2}}\theta\theta\partial_\mu\psi_b\sigma^\mu\bar\theta+\frac{1}{4}\theta\theta\bar\theta\bar\theta\Box \phi_b,\\
\bar{\boldsymbol{\phi}}^b&=\phi^{\ast b}+\sqrt{2}\bar\theta\bar\psi^b+\bar\theta\bar\theta F^{\ast b}
-i\theta\sigma^\mu\bar\theta\partial_\mu\phi^{\ast b}
+\frac{i}{\sqrt{2}}\bar\theta\bar\theta\theta\sigma^\mu\partial_\mu\bar\psi^b+\frac{1}{4}\theta\theta\bar\theta\bar\theta\Box \phi^{\ast b},\\
\tilde{\boldsymbol{\phi}}^b&=\tilde\phi^b+\sqrt{2}\theta\tilde\psi^b+\theta\theta \tilde F^b+i\theta\sigma^\mu\bar\theta\partial_\mu\tilde\phi^b
-\frac{i}{\sqrt{2}}\theta\theta\partial_\mu\tilde\psi^b\sigma^\mu\bar\theta+\frac{1}{4}\theta\theta\bar\theta\bar\theta\Box \tilde\phi^b,\\
\bar{\tilde{\!\!\boldsymbol{\phi}}}_b&=\tilde\phi^\ast_b+\sqrt{2}\bar\theta\bar{\tilde\psi}_b
+\bar\theta\bar\theta \tilde F_b^\ast-i\theta\sigma^\mu\bar\theta\partial_\mu\tilde\phi_b^\ast
+\frac{i}{\sqrt{2}}\bar\theta\bar\theta\theta\sigma^\mu\partial_\mu\bar{\tilde\psi}_b
+\frac{1}{4}\theta\theta\bar\theta\bar\theta\Box\tilde\phi^\ast_b,
\end{split}
\ee
which are fundamental (denoted by subscript Roman letters) and anti-fundamental
(denoted by supscript Roman letters) representations, respectively, under a $SU(N)$ global symmetry
while a Goldstino multiplet $X$ which is a singlet under the symmetry~\footnote{As to the conventions
of the superspace formalism we will closely follow those of Ref.~\cite{Wess:1992cp} throughout the article.}.
The Lagrangian for the chiral superfields reads as follows: 
\be\label{eq:lagwogauge}
{\cal L}_{l\sigma}=
\int \mathrm{d}^2\theta \mathrm{d}^2\bar\theta \big(\,\bar{\!\boldsymbol{\phi}}^b\boldsymbol\phi_b
+\boldsymbol{\tilde\phi}^b\,\, \bar{\tilde{\!\!\boldsymbol\phi}}_b+\bar{X} X\big)
+\bigg[\int \mathrm{d}^2\theta X\boldsymbol{\tilde\phi}^b\boldsymbol\phi_b+h.c.\bigg],
\ee
where we omit additional superpotential for $X$, and K\"{a}hler potential and superpotential
of the hidden sector where supersymmetry is spontaneously broken. 

The K\"{a}hler potential and superpotential as in eq.~(\ref{eq:lagwogauge}) are independently 
invariant under an infinitesimal global transformation:
\be
\begin{aligned}
\delta\boldsymbol\phi_b = i\epsilon_a T^{\boldsymbol{a}\,c}_{\,\,b}\boldsymbol\phi_c,&\quad
\delta\,\bar{\!\boldsymbol{\phi}}^b=-i\epsilon_a \,\bar{\!\boldsymbol{\phi}}^c T^{\boldsymbol{a}\,b}_{\,\,c},\\
\delta\boldsymbol{\tilde\phi}^b=-i\epsilon_a\boldsymbol{\tilde\phi}^c T^{\boldsymbol{a}\,b}_{\,\,c},&\quad
\delta\,\,\bar{\tilde{\!\!\boldsymbol{\phi}}}_b=i\epsilon_a T^{\boldsymbol{a}\,c}_{\,\,b}\,\,\bar{\tilde{\!\!\boldsymbol{\phi}}}_c,\\
\delta X=0.&
\end{aligned}
\ee
where $\epsilon^a$ is a real infinitesimal constant and  $T^{\boldsymbol{a}}$ is a $SU(N)$ generator. 
Moreover, the superpotential is also invariant under an {\it extended} global transformation:
\be
\begin{aligned}\label{eq:gent}
\delta \boldsymbol\phi_b=i\epsilon\Lambda_a T^{\boldsymbol{a}\,c}_{\,\,b}\boldsymbol\phi_c,&\quad
\delta\,\bar{\!\boldsymbol{\phi}}^b= -i\epsilon\bar\Lambda_a\,\bar{\!\boldsymbol{\phi}}^c T^{\boldsymbol{a}\,b}_{\,\,c},\\
\delta \boldsymbol{\tilde\phi}^b=-i\epsilon\Lambda_a \boldsymbol{\tilde\phi}^c T^{\boldsymbol{a}\,b}_{\,\,c},&\quad
\delta\,\,\bar{\tilde{\!\!\boldsymbol{\phi}}}_b= i\epsilon \bar\Lambda_aT^{\boldsymbol{a}\,c}_{\,\,b}\,\,\bar{\tilde{\!\!\boldsymbol{\phi}}}_c,\\
\delta X =0.&
\end{aligned}
\ee
with $\Lambda_a(x,\theta,\bar\theta)$ a chiral superfield. 
On the other hand, the K\"{a}hler potential are {\it not} in general invariant under these transformations,
because $\Lambda_a \neq\bar\Lambda_a$.
For general fields the change in the action must therefore be of the form
\be\label{eq:action}
\delta I = i\epsilon\int \mathrm{d}^4x \int \mathrm{d}^2\theta \mathrm{d}^2\bar\theta(\Lambda_a-\bar\Lambda_a) {\cal J}^a,
\ee
where
\be\label{eq:sc}
{\cal J}^a(x,\theta,\bar\theta)\equiv
\,\,\bar{\!\boldsymbol\phi}T^a\boldsymbol\phi-\boldsymbol{\tilde\phi} T^a\,\,\bar{\tilde{\!\!\boldsymbol\phi}}.
\ee
If the equations of motion are satisfied
then the action (\ref{eq:action}) is stationary under {\it any} variation in the superfields,
so the integral must vanish for any chiral superfield $\Lambda_a(x,\theta,\bar\theta)$.
This implies that ${\cal J}$, which is called {\it the current superfield}~\cite{Salam:1976ib},
must meet the following conditions:
\be
\bar D^2 {\cal J}^a=D^2 {\cal J}^a=0,
\ee
and is of the form\footnote{It is called a {\it linear} superfield.}
\begin{multline}\label{eq:expj}
{\cal J}^a(x,\theta,\bar\theta)=J^a(x)+i\theta j^a(x) -i\bar\theta\bar j^a(x)
-\theta \sigma^\mu \bar \theta j^a_\mu(x)\\
+\frac{1}{2}\theta\theta\bar\theta\bar\sigma^\mu \partial_\mu j^a(x)
-\frac{1}{2}\bar\theta\bar\theta\theta\sigma^\mu\partial_\mu\bar j^a(x)
+\frac{1}{4}\theta\theta\bar\theta\bar\theta \Box J^a(x),
\end{multline}
with the conserved {\it vector} current associated with the $SU(N)$ global symmetry,
$\partial^\mu j^a_\mu=0.$

The components of the current supefield are straightforwardly read
by comparing eq.~(\ref{eq:sc}) with eq.~(\ref{eq:expj}):
\be\label{eq:expcur}
\begin{aligned}
J^a&= \phi^\dagger T^a\phi-\tilde{\phi}T^a\tilde\phi^\dagger,\\
j^a&=-\sqrt{2}i(\phi^\dagger T^a \psi-\tilde\psi T^a\tilde\phi^\dagger),\\
\bar j^a&=\sqrt{2}i(\bar\psi T^a \phi-\tilde\phi T^a \bar{\tilde\psi}),\\
j_\mu^a&=-i\phi^\dagger T^a\overleftrightarrow{\partial_\mu} \phi
+i\tilde\phi T^a\overleftrightarrow{\partial_\mu} \tilde\phi^\dagger
-\bar\psi\bar\sigma_\mu T^a\psi - \tilde\psi\sigma_\mu T^a\bar{\tilde\psi},
\end{aligned}
\ee
where $\overleftrightarrow{\partial}\equiv-\overleftarrow{\partial}+\overrightarrow{\partial}$.
Notice that the $F$-components of the messengers do not appear in eq.~(\ref{eq:expcur}),
and each component is decomposed into the two parts - one is composed only of $\boldsymbol\phi$
and the other is composed only of $\boldsymbol{\tilde\phi}$. 
The vector current $j_\mu^a$ is the {\it Noether current} associated with the continuous transformations~(\ref{eq:gent})
so the corresponding Noether charge is $Q^a=\int \mathrm{d}^3\mathbf{x} j^a_0(\mathbf{x},t)$.
From (\ref{eq:expcur}) we easily identify  $j_\mu^a$ with a vector current of {\it the linear sigma model}.
That is, the Lagrangian (\ref{eq:lagwogauge}) is the (ungauged) {\it supersymmetric linear sigma model}.

We further decompose $j^a_\mu$ into the bosonic and fermionic segments as follows,
\be
\begin{aligned}
j_\mu^{aB}&=-i\phi^\dagger T^a\overleftrightarrow{\partial_\mu} \phi
+i\tilde\phi T^a\overleftrightarrow{\partial_\mu} \tilde\phi^\dagger,\\
j_\mu^{aF}&=-\bar\psi\bar\sigma_\mu T^a\psi - \tilde\psi\sigma_\mu T^a\bar{\tilde\psi},
\end{aligned}
\ee 
and then examine the 3-point function of the fermionic vector current,
\be \langle j_\mu^{aF}(x_1) j_\nu^{bF}(x_2) j_\rho^{cF}(x_3)\rangle\propto \mbox{Tr}\,[T^aT^bT^c]. \ee
For $SU(N)$ symmetry group with $N\ge 3$,  each one-loop triangle diagram composed of $\psi$
and $\tilde\psi$, respectively, does not vanish but the sum of the two diagrams does vanish~\footnote{For
$SU(2)$ symmetry group, it automatically vanishes due to the absence of totally symmetric structure constant.}.
This feature is indispensible when the global symmetry is gauged.
In other words, a model is consistent if there is no gauge anomaly.
The same is true for the 3-point functions of the bosonic current,
$\langle j_\mu^{aB}(x_1) j_\nu^{aB}(x_2) j_\rho^{aB}(x_3)\rangle$.
One can also take into account various 3-point functions of $J^a, j^a$ and $\bar j^a$.
But there are no such restrictions like no gauge anomaly so these 3-point functions are generally
nonzero.

Using canonical (anti-)commutation relations for $(\phi,\phi^\ast),(\tilde\phi,\tilde\phi^\ast),(\psi,\bar\psi)$,
and $(\tilde\psi,\bar{\tilde\psi})$, we can check that for the time components of the $j^a_\mu$
the local current algebra remains valid~\cite{deAzcarraga:1986yq} 
\be[j^a_0(x),j^b_0(y)]_{x^0=y^0}=it^{abc}j^c_0\delta^{(3)}(\mathbf{x-y}), \ee
where the structure constants $t^{abc}$ is defined as $[T^a, T^b]=it^{abc}T^c$.

Though $\langle X\rangle$ gets a nonzero VEV
the Lagrangian (\ref{eq:lagwogauge}) respects the $SU(N)$ global symmetry
so that (\ref{eq:expcur}) still holds true. 
On top of that, so does (\ref{eq:expcur}) even in case that  $\langle F_X\rangle \neq 0$.
The essential idea of these arguments is the following.
{\it The use of current superfield is justified
so long as supersymmetry is spontaneously broken and the global vector symmetry is conserved.}
In the next section, we will weakly gauge this global vector symmetry
and then identify it with the SM gauge symmetries.

In order to employ the current superfield for the study of gauge mediation
we must  set a restriction that all the messenger fields are massive, {\it i.e.}
no tachyonic mode in the sfermions of the MSSM.
For example, in our model this is met if $|\langle X\rangle |^2 > \langle F_X\rangle$. 
A nonzero $\langle X\rangle$ in GGM is equivalent to explicit breakdown of the axial vector symmetry
in the sigma model while a nonzero $\langle F_X\rangle$ correponds to soft breakdown of the axial vector symmetry.
In general the axial vector symmetry of the linear sigma model is broken by quantum effects.
It is realistic that explicit breaking effect is, if any, larger than soft breaking one in the linear sigma model.
Thus we can fathom why the condition $|\langle X\rangle |^2 > \langle F_X\rangle$ is met in the linear sigma model.

Now we turn to an Abelian symmetry associated with supersymmetry, so called the $R$ symmetry
which must be spontaneously broken in any realistic model.
If the $X$ is charged under the $R$ symmetry then the $R$ symmetry is spontaneously broken 
or can be anomalous by quantum effects.
That is, the 3-point correlator of 1 $R$- and 2 local vector currents in general does not vanish,
\be
\sum Q_i^2R\propto \sum R \neq 0,
\ee
where $Q_i$ is the $U(1)$ charge of the messenger fields in a triangle loop.
In this regard, GGM can inherently embrace the $R$-axion physics.
On the other hand, if the $X$ is neutral under the $R$ symmetry
then the $R$ symmetry should be broken in the hidden sector.

\subsection{Current superfield and supersymmetry currents}
In this section we shall obtain the algebra among the components of the current superfield,
and the supercharges for the Lagrangian~(\ref{eq:lagwogauge}).
In supersymmetric quantum field theories  the supercharges ${\cal Q}_\alpha$ and $\overline{\cal Q}_{\dot\alpha}$
may be represented in terms of supersymmetry currents $S^\mu_\alpha$ and $\overline{S}^\mu_{\dot\alpha}$,
respectively:
\be
\begin{aligned} 
{\cal Q}_\alpha &= \int \mathrm{d}^3\mathbf{x} S^{\,0}_\alpha,\\
\bar{\cal Q}_{\dot\alpha} &= \int \mathrm{d}^3\mathbf{x} \bar{S}^{\,0}_{\dot\alpha}.
\end{aligned}
\ee
The supersymmetry currents for the Lagrangian (\ref{eq:lagwogauge}) are given by,
\be
\begin{aligned}
\frac{S^{\,\mu}_\alpha}{\sqrt{2}}&= 
(\sigma^\nu\bar\sigma^\mu\psi)^T_\alpha\partial_\nu \phi^\ast
+(\sigma^\nu\bar\sigma^\mu\tilde\psi)^T_\alpha\partial_\nu \tilde\phi^\ast
+\cdots,\\
\frac{\bar{S}^{\,\mu}_{\dot\alpha}}{\sqrt{2}}&= 
(\bar\psi\bar\sigma^\mu\sigma^\nu)_{\dot\alpha}\partial_\nu\phi
+(\bar{\tilde\psi}\bar\sigma^\mu\sigma^\nu)_{\dot\alpha}\partial_\nu\tilde\phi
+\cdots,
\end{aligned}
\ee
where we omitted the other terms irrelevant to the evaluation.

Using the equal-time $(x^0=y^0)$ commutation relations, 
\begin{gather}
[\phi_b(x),\phi_c(y)]=[\phi^{\ast b}(x),\phi^{\ast c}(y)]=
[ \tilde{\phi}^b(x),\tilde{\phi}^c(y)] =[\tilde\phi^\ast_b(x),\tilde\phi^\ast_c(y)]=0,\\
[\phi_b(x),\partial_0 \phi^{\ast c}(y)]
=[\tilde\phi^c(x),\partial_0 \tilde\phi^\ast_b(y)]=
i\delta^{(3)}(\mathbf{x-y})\delta_b^c,
\end{gather}
We verify that the commutator of ${\cal Q}_\alpha$ and $J^a$ yields
the spinor component of the current superfield
\bea\label{eq:comm1}
[{\cal Q}_\alpha, J^a(0)]&=&\int \mathrm{d}^3\mathbf{x} [S^{\,0}_\alpha(x),J^a(0)]\nonumber\\
&=&\sqrt{2} \int \mathrm{d}^3\mathbf{x}\big(\psi_{\alpha b}[\partial_0\phi^{\ast b}(x),\phi_d(0)]\phi^{\ast c} T^{\boldsymbol{a}\,d}_{\,\,c}
-\tilde\psi_\alpha^b[\partial_0\tilde\phi^\ast_b (x),\tilde\phi^c(0)] T^{\boldsymbol{a}\,d}_{\,\,c}\tilde\phi^\ast_d\big)\nonumber\\
&=&\sqrt{2} \int \mathrm{d}^3\mathbf{x}\big(\phi^{\ast b}T^{\boldsymbol{a}\,c}_b\psi_{\alpha c}
-\tilde\psi_\alpha^b T^{\boldsymbol{a}\,c}_{\,\,b} \tilde\phi^\ast_c\big)(-i)\delta^{(3)}(\mathbf{x})\nonumber\\
&=&-\sqrt{2}i(\phi^\dagger T^{\boldsymbol{a}} \psi_\alpha-\tilde\psi_\alpha T^{\boldsymbol{a}}\tilde\phi^\dagger)(0)\nonumber\\
&=&j^a_\alpha(0).
\eea
Likewise, we find the commutation relation between $\overline{\cal Q}_{\dot\alpha}$ and $J^a$,
\be\label{eq:comm2}
[\bar{{\cal Q}}_{\dot\alpha},J^a(0)]=\bar j^a_{\dot\alpha}(0).
\ee
Furthermore, following from the equal-time $(x^0=y^0)$ anti-commutation relations,
\begin{gather}
\{\psi_b(x),\psi_c(y)\}=\{\bar\psi^b(x),\bar\psi^c(y)\}=
\{\tilde{\psi}^b(x),\tilde{\psi}^c(y)\} =\{\bar{\tilde\psi}_b(x),\bar{\tilde\psi}_c(y)\}=0,\\
\{\psi_b(x),\bar\psi^c(y)\}
=\{\tilde\psi^c(x),\bar{\tilde\psi}_b(y)\}=\delta^{(3)}(\mathbf{x-y})\delta_b^c,
\end{gather}
we also find the vector component of the current superfield,
\be\label{eq:comm3}
j^a_\mu = -\frac{1}{4}\bar\sigma^{\dot\alpha\alpha}_\mu 
(\{\bar{\cal Q}_{\dot\alpha},[{\cal Q}_\alpha,J^a]\}
-\{{\cal Q}_\alpha,[\bar{\cal Q}_{\dot\alpha},J^a]\}),
\ee
and the commutation relations
\be\label{eq:comm4}
\{{\cal Q}_\alpha,[{\cal Q}_\beta,J^a]\}
=\{\bar{\cal Q}_{\dot\alpha},[\bar{\cal Q}_{\dot\beta},J^a]\}=0.
\ee

These (anti-)commutation relations (\ref{eq:comm1}), (\ref{eq:comm2}), (\ref{eq:comm3}), and  (\ref{eq:comm4})
are a direct consequence both of (spontaneously broken) supersymmetry,
and of the conservation of the global vector symmetry.
Once again the latter condition is entailed to gauge the global vector symmetry
and in turn to identify it with the SM gauge symmetry.
Therefore, so long as the two conditions are met
the same (anti-)commutation relations are applied to more complex models
- multiple linear sigma models and non-linear sigma models.
In fact, the usefulness of these (anti)-commutation relations is to evaluate the other components
from the scalar component of a current superfield in the non-linear sigma model.  

\section{Current correlators}\label{sect:cur}
In this section we consider correlation functions of the current superfield in the linear sigma model.
Our presumption is the existence of the current superfield along with spontaneous supersymmetry breaking,
which indicates that the degrees of freedom for scalars are equal to that of fermions,
and the scalar masses are split from their erstwhile supersymmetric values, while the fermions are not.
We need to compute multipoint correlation functions of the current superfield in which
the intermediate states (or propagators) are particularized by their masses.
For simplicity we consider only linear sigma models with a single vectorlike messenger,
$\boldsymbol{\phi}$ and $\tilde{\boldsymbol{\phi}}$. The generalization with multiple messengers
is straightforward.

We first convert the scalar global symmetry group eigenstates $(\phi,\tilde\phi)$
into the mass eigenstates $(\phi_+,\phi_-)$:
\be
\left(\begin{array}{c} \phi_- \\ \phi_+\end{array}\right) =
\frac{1}{\sqrt{2}}\left(\begin{array}{cc} 1 & 1\\ 1& -1\end{array}\right)
\left(\begin{array}{c} \phi \\ \tilde\phi^\dagger\end{array}\right),
\ee
The mass of the scalar field $\phi_+(\phi_-)$ is $M_+(M_-)$.
On the other hand, the two messenger fermions $\psi$ and $\tilde\psi$ have the same mass as $M_0$.
Note that $M^2_\pm=M^2_0\pm F$ and $M_0=\langle X\rangle, F=\langle F_X\rangle$ in (\ref{eq:lagwogauge}).   
Then the current superfield as in (\ref{eq:expcur}) is expressed in terms of mass eigenstates:
\be\label{eq:expcur2}
\begin{aligned}
J^a&= \phi^\dagger_+ T^a\phi_-+\phi^\dagger_- T^a\phi_+,\\
j^a&=-i[(\phi^\dagger_+ +\phi^\dagger_-) T^a \psi +\tilde\psi T^a (\phi_+-\phi_-)],\\
\bar j^a&=i[(\bar \psi T^a(\phi_+ +\phi_-)+(\phi^\dagger_+ - \phi^\dagger_-)T^a\bar{\tilde\psi}]),\\
j_\mu^a&=-i\phi^\dagger_+ T^a\overleftrightarrow{\partial_\mu} \phi_-
-i\phi^\dagger_- T^a\overleftrightarrow{\partial_\mu} \phi_+
-\bar\psi\bar\sigma_\mu T^a\psi - \tilde\psi\sigma_\mu T^a\bar{\tilde\psi}.
\end{aligned}
\ee
Notice that $j^a(\bar j^a)$ contains the two linear combinations of $\phi_+$ and $\phi_-$.
One is a ``+" sign linear combination and the other is a ``-" sign linear combination.
On the other hand, $J^a$ and $j^a_\mu$ contain a product of $\phi_+$ and $\phi_-$.

We now pursue general features of the multipoint correlation functions in the linear sigma model
in Section~\ref{sect2}.
We first consider the 1-point function derived from an Abelian global symmetry.
Due to current conservation and Lorentz invariance the only nonzero 1-point function can be 
\be \langle J(x)\rangle=\langle J(0)\rangle=\zeta\neq 0,\ee
implying that the scalar component of the messenger chiral superfield has a nonzero VEV.
The Abelian global symmetry is then spontaneously broken
so the corresponding Nambu-Goldstone boson(NGB) emerges.
Once the Abelian symmetry is gauged in the framework of GGM
the NGB is eaten by the Abelian gauge field so that the Abelian gauge symmetry is broken.
However we should have the SM $U(1)_Y$ gauge symmetry unbroken before electroweak symmetry breaking
so the 1-point function is {\it not} allowed~\footnote{At this point we must be cautious to draw this conclusion.
Here we assume that breaking of the Abelian symmetry does not occur at the same scale as electroweak
symmetry breaking does.}.

Now we turn to the 2-point functions of the current superfield.
There are only four nonzero 2-point functions built from the current superfield~\cite{Meade:2008wd}.
They are expressed in terms of the free field propagators as follows (See Appendix~\ref{app1}):
\bea
\langle J^a(x)J^a(0)\rangle &=&-2 \mathbf{C}(\mathrm{R}) \Delta_F(x;M_+) \Delta_F(x;M_-),\label{eq:220}\\
\langle j^a_\alpha(x)\bar j^a_{\dot\alpha}(0)\rangle &=& 2i\mathbf{C}(\mathrm{R})\big[\Delta_F(x;M_+)+\Delta_F(x;M_-)\big]
\sigma_{\alpha\dot\alpha}^\mu \partial_\mu\Delta_F(x;M_0),\label{eq:22x1}\\
\langle j^a_\alpha(x) j^a_\beta(0)\rangle
&=&2 \mathbf{C}(\mathrm{R})[\Delta_F(x;M_+)-\Delta_F(x;M_-)]\varepsilon_{\alpha\beta}M_0\Delta_F(x;M_0),\label{eq:2212}\\
\langle j^a_\mu (x) j^a_\nu (0)\rangle
&=& 2 \mathbf{C}(\mathrm{R})\big[\Delta_F(x;M_-)\partial_\mu\partial_\nu \Delta_F(x;M_+)
-\partial_\mu\Delta_F(x;M_+)\partial_\nu\Delta_F(x;M_-)\nonumber\\
&&\qquad\qquad+\Delta_F(x;M_+)\partial_\mu\partial_\nu \Delta_F(x;M_-)
- \partial_\mu\Delta_F(x;M_-)\partial_\nu\Delta_F(x;M_+)\nonumber\\
&&\qquad\qquad+2\eta_{\mu\nu}\big(-\partial_\rho \Delta_F(x;M_0)\partial^\rho \Delta_F(x;M_0)
+M_0^2\Delta^2_F(x;M_0)\big) \nonumber\\
&&\qquad\qquad+ 4\partial_\mu \Delta_F(x;M_0)\partial_\nu \Delta_F(x;M_0) \big]\label{eq:22z1},
\eea
where $\Delta_F(x;M)$ is the propagator for a scalar field with mass $M$,
\be \Delta_F(x;M) =\frac{1}{\Box-M^2}, \ee
and  $\mathbf{C}(\mathrm{R})$ is the quadratic invariant of the fundamental representation under a $SU(N)$ global symmetry group,
\be \mbox{Tr} [T^aT^b] =   \mathbf{C}(\mathrm{R})\delta^{ab}. \ee

We make several comments on the explicit expression for the 2-point function.
First of all, all the four 2-point functions are given as a sum of product of the two propagators.
In particular the scalar 2-point function is the product of the two scalar propagators. 
Comparing the first spinor 2-point function with the second spinor 2-point function, we find that
the first spinor 2-point function contains a ``+" sign linear combination of the two scalar propagators
while the second spinor 2-point function does a ``-" sign linear combination of the two scalar propagators.
Moreover, the former contains a helicity-preserving fermion propagator
while the latter does a helicity-flipping fermion propagator.
The correlation between the helicity in the fermion propagator
and the sign of the linear combination of the two scalar propagators
stems from the structure of the spinor current as in (\ref{eq:expcur2}).

In supersymmetric limit ($M_+=M_-=M_0$) the 2-point functions can be written as
\bea
\langle J^a(x)J^a(0) \rangle &=& -2\mathbf{C}(\mathrm{R}) \Delta^2_F(x;M_0),\\
\langle j^a_\alpha(x)\bar j^a_{\dot\alpha}(0)\rangle
&=&-i\sigma^\mu_{\alpha\dot\alpha}\partial_\mu \langle J^a(x)J^a(0)\rangle, \label{eq:15exp}\\
\langle j^a_\alpha(x) j^a_\beta(0)\rangle&=&0, \label{eq:12exp}\\
\langle j^a_\mu(x)j^a_\nu(0)\rangle &=&
 (\eta_{\mu\nu}\partial_\rho\partial^\rho-\partial_\mu\partial_\nu) \langle J^a(x)J^a(0)\rangle
-2\eta_{\mu\nu}\mathbf{C}(\mathrm{R}) \Delta_F(x;M_0).\label{eq:25exp}
\eea
The existence of the last term on the right-hand  side as in eq.~(\ref{eq:25exp}) makes the relation look pointless.
However, once the global symmetry is gauged it is cancelled by additional matrix elements
introduced for gauge invariance.

Finally we make a brief comment on the relations among the 2-point functions.
For the case that supersymmetry is spontaneously broken there are no simple relations
like eq.~(\ref{eq:15exp}), (\ref{eq:12exp}) and (\ref{eq:25exp}).

\section{Gauging $SU(N)$ global symmetry}
We have so far discussed a $SU(N)$ global symmetry in Wess-Zumino model.
Now we weakly gauge the $SU(N)$ global symmetry such that the Lagrangian
as in (\ref{eq:lagwogauge}) is transformed into 
\be
{\cal L}=\int \mathrm{d}^2\theta \mathrm{d}^2\bar\theta
(\,\bar{\!\boldsymbol\phi}\, e^{2gV}\!\boldsymbol\phi
+\boldsymbol{\tilde\phi}\, e^{-2gV}\,\bar{\tilde{\!\!\boldsymbol\phi}}
+\bar XX)
+\bigg[\int \mathrm{d}^2\theta\bigg( \frac{1}{4}W^\alpha W_\alpha
+X\boldsymbol{\tilde\phi}\boldsymbol\phi\bigg)+ h.c.\bigg],
\ee
which is invariant under the supersymmetric gauge transformations:
\be
\begin{split}
\boldsymbol\phi&\to e^{-2ig\Lambda}\boldsymbol\phi,\\
\boldsymbol{\tilde\phi}&\to \boldsymbol{\tilde\phi}\,e^{2ig\Lambda},\\
e^{2gV}&\to e^{-2ig\bar\Lambda}e^{2gV}e^{2ig\Lambda},
\end{split}
\ee
where $g$ is the gauge coupling, and $\Lambda=\Lambda_a T^a,\bar\Lambda=\bar\Lambda_a T^a$
are matrices composed of complex functions of $(x,\theta,\bar\theta)$ characterized by the conditions
\be
\bar{D}_{\dot\alpha}\Lambda^a =0,\quad D_\alpha \bar\Lambda^a =0.
\ee
In the Wess-Zumino gauge, one can write the off-shell Lagrangian in terms of the component fields as
\be\label{eq:wzlag}
\begin{aligned}
{\cal L}&= -\frac{1}{4}v^a_{\mu\nu}v^{a\mu\nu}-i\lambda^a\sigma^\mu{\cal D}_\mu\bar\lambda^a
+\frac{1}{2}D^aD^a\\
&\quad-{\cal D}^\mu \phi^\dagger {\cal D}_\mu \phi-i\bar\psi\bar\sigma^\mu{\cal D}_\mu\psi+F^\dagger F
-{\cal D}^\mu \tilde\phi {\cal D}_\mu \tilde\phi^\dagger-i\tilde\psi\sigma^\mu{\cal D}_\mu\bar{\tilde\psi}
+\tilde F \tilde F^\dagger\\
&\quad+\sqrt{2}ig(\phi^\dagger T^a\psi - \tilde \psi T^a \tilde\phi^\dagger)\lambda^a
-\sqrt{2}ig\bar\lambda^a(\bar\psi T^a \phi - \tilde\phi T^a\bar{\tilde\psi})\\
&\quad+gD^a(\phi^\dagger T^a\phi-\tilde \phi T^a\tilde \phi^\dagger),
\end{aligned}
\ee
where
\be
\begin{aligned}
{\cal D}_\mu \phi&= \partial_\mu \phi+igv^a_\mu T^a \phi,\\
{\cal D}_\mu \tilde \phi&= \partial_\mu \tilde \phi-igv^a_\mu T^a \tilde \phi,\\
{\cal D}_\mu \psi &= \partial_\mu\psi+igv^a_\mu T^a\psi,\\
{\cal D}_\mu \tilde\psi &= \partial_\mu\tilde\psi-igv^a_\mu T^a\tilde\psi,\\
{\cal D}_\mu \lambda^a&= \partial_\mu\lambda^a-gt^{abc}v^b_\mu\lambda^c,\\
v^a_{\mu\nu}&=\partial_\mu v^a_\nu-\partial_\nu v^a_\mu-gt^{abc}v^b_\mu v^c_\nu.
\end{aligned}
\ee

Promoting the global symmetry to a local symmetry
yields coupling of the globally symmetric theory to non-Abelian gauge vector superfield
to first order in $g$,
\be
{\cal L}'=- 2g \int \mathrm{d}^2\theta\mathrm{d}^2\bar\theta{\cal J}^a V^a=
g(-v^a_\mu j^{a\mu}-j^a\lambda^a-\bar\lambda^a\bar j^a+  D^a J^a).
\ee

Now let us consider the case in which two external gauge bosons, gauginos or $D$ components
of the vector superfield attach to an internal messenger loop.
The internal messenger loop is anything but the 2-point functions of the current superfield in Section~\ref{sect:cur}.
But we should be careful if the two external lines are gauge bosons.
We must include contributions arising from the terms of order $g^2$ in (\ref{eq:wzlag}) for gauge invariance,
\be\label{eq:2ndg}
{\cal L}{''}=-\frac{g^2}{2}\mathbf{C}(\mathrm{R}) v^a_\mu v^\mu_a (\phi^\dagger \phi +\tilde\phi \tilde\phi^\dagger).
\ee
Note that $\phi^\dagger \phi +\tilde\phi \tilde\phi^\dagger$ is the same as the ungauged K\"{a}hler potential 
except that it is made not of chiral superfields but of its scalar components.
The role of the term in eq.~(\ref{eq:2ndg}) is to maintain the Ward identity
such that the tensor structure of vacuum polarization of the gauge boson $i\Pi_{\mu\nu}(p^2)$ is proportional to
the projector $(\eta_{\mu\nu} -p_\mu p_\nu/p^2)$. We will discuss it more in the next section.

\section{Regularizating the current correlators}\label{sec:loop}
In this section the calculations of the four 2-point functions
(\ref{eq:220}), (\ref{eq:22x1}),  (\ref{eq:2212}) and  (\ref{eq:22z1}) are carried out at one loop,
using dimensional regulaization.
It turns out that they are the messenger contributions to the 2-point functions for the component fields
of the vector superfield.
We now switch from a {\it Euclidean} to a {\it Minkowski} metric.
\subsection{2-point function for $D$ field}
\FIGURE{\epsfig{file=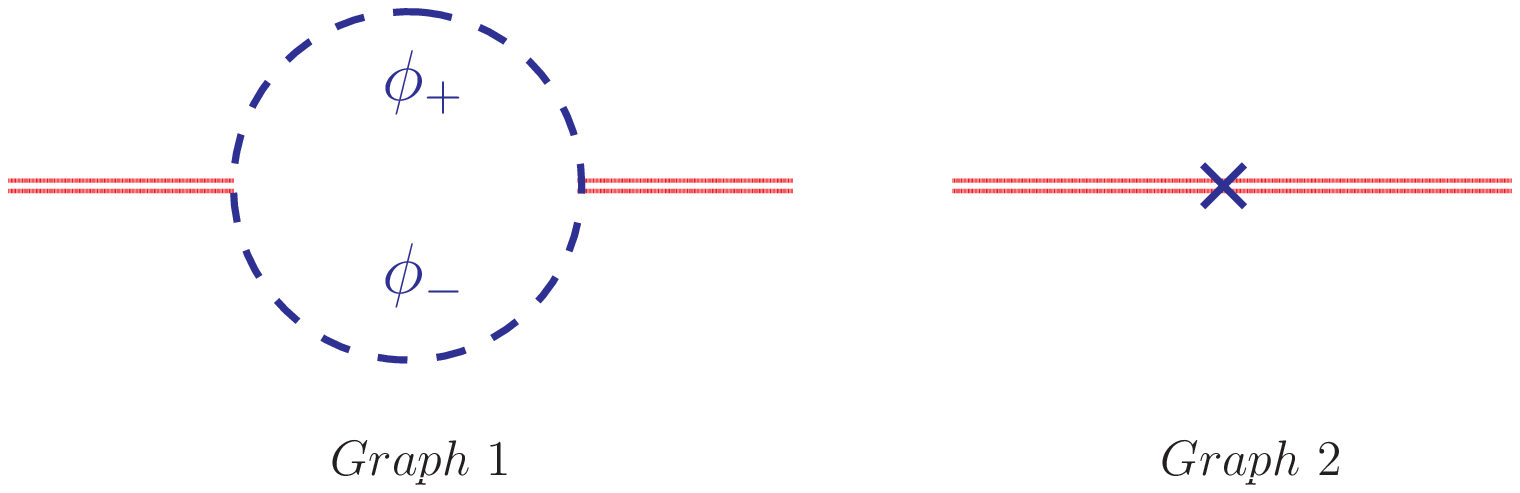, width=10 cm}
\caption{The one-loop and counterterm corrections to the $D^2$-term.
The blue dotted lines in {\it Graph} 1 denote the messenger scalar fields
in mass eigenstate while the cross in {\it Graph} 2 the counterterm vertex.}\label{fig:omega1}}
We begin with the 2-point function (\ref{eq:220}) which turns outs to be one loop correction
to the $D$ field propagator.
The graph 1 in Figure~\ref{fig:omega1} is the contribution from the scalar 2-point function of the current superfield
while the graph 2 is the counterterm.
Using dimensional regularization the amplitude for Graph 1 is given by
\bea
i\Omega(p^2) 
&=& 2g^2 \mathbf{C}(\mathrm{R}) \mu^{2\varepsilon}\int \frac{d^d q}{(2\pi)^d}
\frac{1}{(q^2-M^2_+)[(q+p)^2-M^2_-]}\\\nonumber
&=& \frac{2ig^2 \mathbf{C}(\mathrm{R})}{16 \pi^2}\bigg\{\frac{1}{\varepsilon}-\gamma+\log 4\pi
+\int^1_0dy\log\bigg[\frac{\mu^2}{M^2_0+(1-2y)F-y(1-y)p^2}\bigg]+{\cal O}(\varepsilon)\bigg\},
\eea
where $\mu$ is the renormalization scales.~\footnote{Divergent integrals are computed by performing
an analytic continuation to $D=4-2\varepsilon$ spacetime dimensions in the $\overline{MS}$
renormalization scheme.}
This leads to the counterterm
\be\label{eq:rend}
Z_D-1=-\frac{g^2 \mathbf{C}(\mathrm{R})}{16 \pi^2}\frac{2}{\varepsilon}.
\ee
Thus the {\it regularized} scalar 2-point function at one loop is given by
\bea\label{eq:omegar}
\Omega_R(p^2)&=&\frac{2g^2 \mathbf{C}(\mathrm{R})}{16 \pi^2}\bigg\{
\frac{x}{\hat p^2_0}\log\frac{1-x}{1+x}-\frac{1}{2}\log(1-x^2)+\log\frac{\mu^2}{M^2_0}+2\\\nonumber
&&+\bigg(\frac{2x}{\hat p^2_0}-1\bigg)h\bigg(\frac{2x-\hat p^2_0}{\sqrt{4x^2-4\hat p^2_0+\hat p^4_0}}\bigg)
-\bigg(\frac{2x}{\hat p^2_0}+1\bigg) h\bigg(\frac{2x+\hat p^2_0}{\sqrt{4x^2-4\hat p^2_0+\hat p^4_0}}\bigg)\bigg\},
\eea
where $h(z)\equiv\frac{1}{z}\tan^{-1}z$, $\hat p^2_0\equiv\frac{p^2}{M^2_0}$ and $x=\frac{F}{M^2_0}$.

\FIGURE {\epsfig{file=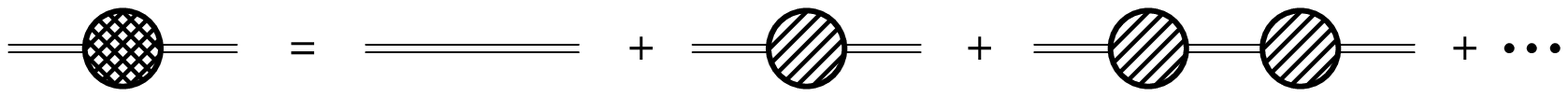, width=16 cm}
\caption{The full expression for $D$ field renormalization.
The line with double-shaded blob denotes the full amplitude for $D^2$,
and the single-shaded blobs the amplitude for the 1PI diagrams
as shown in Figure~\ref{fig:omega1}.}\label{fig:omega2}}
By summing a geometric series of the $D^2$-term and the invariant function $\Omega_R$,
the {\it full} expression for the wavefunction renormalization constant for the $D$ field,
as depicted in fig.~\ref{fig:omega2}, is given by
\be
\frac{1}{1-\Omega_R}=1+\Omega_R+\Omega_R^2+\cdots.
\ee
\subsection{2-point function for gauge field}
We compute the vector 2-point function (\ref{eq:22z1}).
The graphs of Figure~\ref{fig:pi1} denote the messenger one-loops and their counterterm corrections
to the gauge boson propagators.
The graph 1 and 2 have the contributions of the ferminoic and bosonic vector 2-point functions
of the current superfield, respectively.
The graph 3 gives the contribution of the contact term as in eq.~(\ref{eq:2ndg}), and the graph 4 does
the contribution of the counterterm which cancels ultraviolet divergences of the first three graphs,
in particular logarithmic ultraviolet divergences.
Summing up the first three graphs, we have the amplitude for the 1PI graphs (at one loop) as
\be
i\Pi_{\mu\nu}^{[1]}(p^2)= (p_\mu p_\nu-\eta_{\mu\nu}p^2)i\Pi^{[1]}(p^2),
\ee
which is transverse, and the invariant function $i\Pi^{[1]}(p^2)$ is given by 
\bea
i\Pi^{[1]}(p^2) &=& -\frac{2}{3p^2}g^2 \mathbf{C}(\mathrm{R})\mu^{2\varepsilon}\int \frac{d^d q}{(2\pi)^d}\bigg\{
\frac{(p+q)\cdot(p+2q)}{(q^2-M^2_+)[(p+q)^2-M^2_-]}+(M_+\to M_-)\nonumber\\
&&+\frac{4q\cdot(p+q)-8M^2_0}{(q^2-M^2_0)[(p+q)^2-M^2_0]}-\frac{4}{q^2-M^2_+}-\frac{4}{q^2-M^2_-}\bigg\}\nonumber\\
&=& \frac{2ig^2 \mathbf{C}(\mathrm{R})}{16\pi^2}\bigg\{\bigg(\frac{1}{\varepsilon}-\gamma+\log 4\pi\bigg)
+\frac{4}{9}\nonumber\\
&&-\frac{1}{3}\int^1_0 dy (1-y)(1-6y)\log\bigg[\frac{\mu^2}{M^2_+-y(1-y)p^2}\bigg]+(M_+\to M_-)\nonumber\\
&&+4\int^1_0dy y(1-y)\log\bigg[\frac{\mu^2}{M^2_0-y(1-y)p^2}\bigg]\nonumber\\
&&+\frac{4M^2_+}{3p^2} \int^1_0dy \log[1-y(1-y)\frac{p^2}{M^2_+}]
+(M_+\to M_-)+{\cal O}(\varepsilon)\bigg\},
\eea
which leads to the counterterm
\be\label{eq:renv}
(Z_v-1)^{[1]}=-\frac{g^2 \mathbf{C}(\mathrm{R})}{16 \pi^2}\frac{2}{\varepsilon}.
\ee
We note that it is the same as that of $D$ field propagator as in (\ref{eq:rend}), as expected.
\FIGURE {\epsfig{file=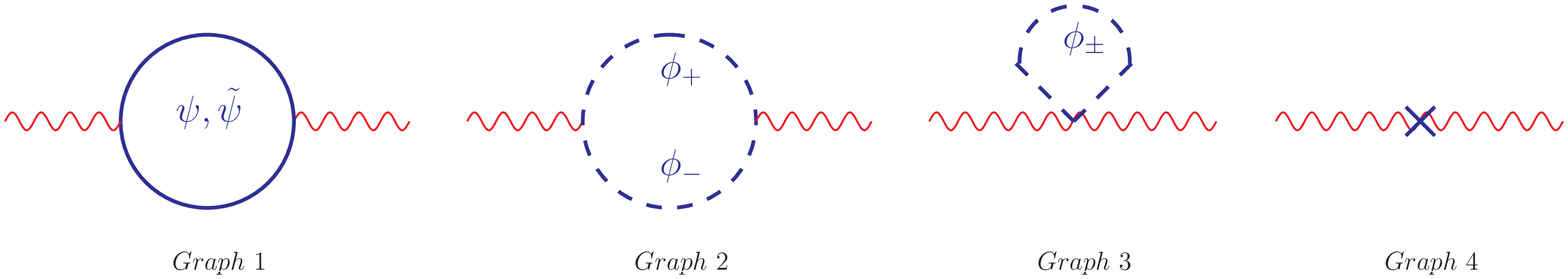, width=15 cm}
\caption{The one-loop and counterterm corrections to a gauge boson propagator.
The blue plain lines in Graph 1 denotes the messenger fermions, the blue dotted lines in Graph
2 and 3 the messenger scalars and the cross in Graph 4 the counterterm vertex.}\label{fig:pi1}}

The {\it regularized} vector 2-point function at one loop level is given by
\bea
\Pi^{[1]}_R(p^2) &=& \frac{2g^2\mathbf{C}(\mathrm{R})}{16 \pi^2}\bigg\{
\frac{1}{6}\log\frac{\mu^2}{M^2_+}+\frac{1}{6}\log\frac{\mu^2}{M^2_-}+
\frac{2}{3}\log\frac{\mu^2}{M^2_0}\nonumber\\
&&-\frac{1}{3}\bigg(\frac{4}{\hat p^2_+}-1\bigg)\bigg[1-h\bigg(\sqrt{\frac{\hat p^2_+}{4-\hat p^2_+}}\bigg)\bigg]
-\frac{1}{3}\bigg(\frac{4}{\hat p^2_-}-1\bigg)\bigg[1-h\bigg(\sqrt{\frac{\hat p^2_-}{4-\hat p^2_-}}\bigg)\bigg]\nonumber\\
&&+\frac{4}{3}\bigg(\frac{2}{\hat p^2_0}+1\bigg)\bigg[1-h\bigg(\sqrt{\frac{\hat p^2_0}{4-\hat p^2_0}}\bigg)\bigg]\bigg\},
\eea
where $\hat p^2_\pm\equiv\frac{p^2}{M^2_\pm}$.
Note that $\Pi^{[1]}_R(p^2)$ goes toward $\Omega_R(p^2)$ in the supersymmetric limit of $x\to 0$.
\FIGURE {\epsfig{file=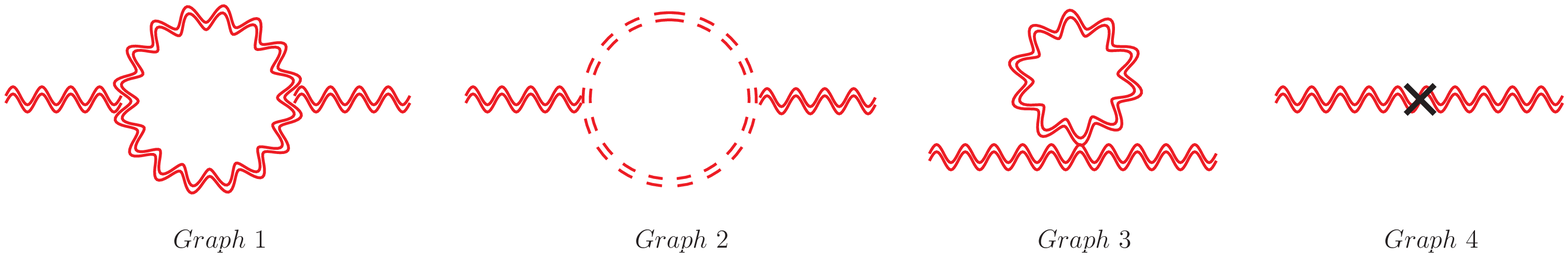, width=15 cm}
\caption{Additional $V$ propagator corrections: all the double lines represent superfield.
Graph 1 denotes the vector loop for its own propagator, Graph 2 the ghost loop, Graph 3 the contacting vector loop,    
and Graph 4 the counterterm.
Graph 3 vanishes in dimensional regularization.}\label{fig:pi2}}

So far we have considered the messenger fields contributions to the 2-point function for the gauge field $v^a_\mu$.
In addition, there are loop contributions from its own field.
Since they are totally independent of supersymmetry breaking, they are the same as those to the gaugino field.
In fact, it is nothing but the contributions from the pure SuperYM theory.
The graphs of fig.~\ref{fig:pi2} yield these contribution to the $V$ propagator.
The contribution of the gauge and gaugino field to the 2-point function for the gauge field is given by
\be
i\Pi^{[2]}(p^2)= -\frac{ig^2\mathbf{C}(\mathbf{G})}{16\pi^2}
\bigg[\bigg(\frac{1}{\varepsilon}-\gamma+\log4\pi\bigg)+\log\frac{\mu^2}{-p^2}+\frac{7}{3}
+{\cal O}(\varepsilon)\bigg],
\ee
which leads to the counterterm
\be
(Z_v-1)^{[2]}=\frac{g^2 \mathbf{C}(\mathbf{G})}{16\pi^2}\frac{1}{\varepsilon}.
\ee
\be
\Pi^{[2]}_R(p^2)=  -\frac{g^2\mathbf{C}(\mathbf{G})}{16\pi^2}
\bigg\{\log\frac{\mu^2}{-p^2}+\frac{7}{3}\bigg\},
\ee
and then the full 2-point function for the gauge field is given by
\be
\Pi_R(p^2) = \Pi^{[1]}_R(p^2)+\Pi^{[2]}_R(p^2).
\ee

By summing a geometric series of a free gauge field propagator and a vacuum polarization tensor
as depicted in fig.~\ref{fig:pi1} and fig.~\ref{fig:pi2},
the {\it full} propagator of the gauge field, as shown in fig.~\ref{fig:pi3}, is given by
\be
\mathbf{\Delta}_{\mu\nu}(p^2)= \frac{-i}{p^2[1-\Pi_R(p^2)]}\bigg(\eta_{\mu\nu}-\frac{p_\mu p_\nu}{p^2}
\bigg)+\frac{-i}{p^2}\bigg(\frac{p_\mu p_\nu}{p^2}\bigg).
\ee
Its longitudinal part gets no corrections, to all orders of perturbation theory~\footnote{We work in Feynman gauge.}.
\FIGURE {\epsfig{file=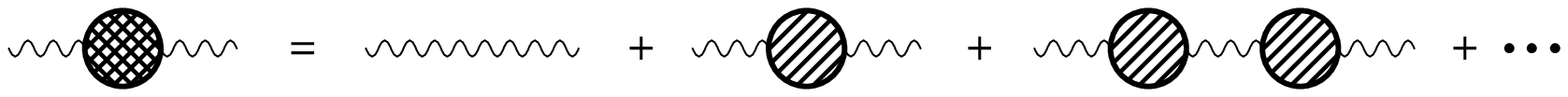, width=15 cm}
\caption{The full propagator for a gauge field.
The line with the double-shaded blob on the left-hand side denotes the full propagator
while the single-shaded blobs the 1PI self energies on the left-hand side.}\label{fig:pi3}}
\subsection{2-point functions for gaugino field}
We compute the spinor 2-point functions (\ref{eq:2212}) and (\ref{eq:22z1}).
There are two different 2-point functions for the  associated with a {\it massive} gaugino field in the basis of 2-component spinors:
one preserves helicity while the other changes it.

To begin with we consider a massless propagator which preserves helicity.
The graph 1 of fig.~\ref{fig:sigma1} is the contribution of the first spinor 2-point function
of the current superfield to the gaugino propagators while the graph 2 is the counterterm
which cancels ultraviolet divergences of the graph 1.
The amplitude for the graph 1 (at one loop) is given by
\be\label{eq:selfenergy}
i\Sigma_{\alpha\dot\alpha}^{[1]}(p^2)=-\sigma^\mu_{\alpha\dot\alpha}p_\mu i\Sigma^{[1]}(p^2),
\ee
where the invariant function $\Sigma^{[1]}(p^2)$ is given by
\bea
i\Sigma^{[1]}(p^2)&=&-\frac{2}{p^2}g^2 \mathbf{C}(\mathrm{R}) \mu^{2\varepsilon}\int \frac{d^d q}{(2\pi)^d}
\frac{1}{[(q+p)^2-M^2_+]}\frac{p\cdot q}{(q^2-M^2_0)}+(M_+\to M_-)\nonumber\\
&=& \frac{2ig^2 \mathbf{C}(\mathrm{R})}{16 \pi^2}\bigg\{\frac{1}{\varepsilon}-\gamma+\log 4\pi
+\int^1_0dxx\log\bigg[\frac{\mu^2}{M^2_0+xF-x(1-x)p^2}\bigg]\nonumber\\
&&+\int^1_0dxx\log\bigg[\frac{\mu^2}{M^2_0-xF-x(1-x)p^2}\bigg]
+{\cal O}(\varepsilon)\bigg\},  
\eea
which leads to the counterterm
\be
(Z_\lambda-1)^{[1]}=-\frac{g^2 \mathbf{C}(\mathrm{R})}{16 \pi^2}\frac{2}{\varepsilon}.
\ee
Once again it is the same as (\ref{eq:rend}) and (\ref{eq:renv}).

The {\it regularized} first spinor 2-point function is then given by 
\bea
\Sigma^{[1]}_R(p^2)&=& \frac{2g^2 \mathbf{C}(\mathrm{R})}{16 \pi^2}\bigg\{
-\frac{1}{2}\bigg(\frac{x}{\hat p^2_0}+1\bigg)^2 h\bigg(\frac{x+\hat p^2_0}{\sqrt{(x+\hat p^2_0)^2-4\hat p^2_0}}\bigg)\nonumber\\
&&-\frac{1}{2}\bigg(\frac{x}{\hat p^2_0}-1\bigg)^2h\bigg(\frac{x-\hat p^2_0}{\sqrt{(x-\hat p^2_0)^2-4\hat p^2_0}}\bigg)\\\nonumber
&&+\frac{1}{2}\bigg(\frac{x^2}{\hat p^4_0}-1\bigg)h\bigg(\frac{x-\hat p^2_0}{\sqrt{(x+\hat p^2_0)^2-4\hat p^2_0}}\bigg)
+\frac{1}{2}\bigg(\frac{x^2}{\hat p^4_0}-1\bigg)h\bigg(\frac{x+\hat p^2_0}{\sqrt{(x-\hat p^2_0)^2-4\hat p^2_0}}\bigg)\\\nonumber
&&+2+\frac{1}{4}\bigg(\frac{x^2}{\hat p^4_0}-\frac{2}{\hat p^2_0}+1\bigg)\log(1-x^2)+\frac{x}{2\hat p^2_0}\log\frac{1-x}{1+x}-\frac{1}{2}\log(1-x^2)
+\log\frac{\mu^2}{M^2_0}\bigg\}.
\eea
As expected, $\Sigma^{[1]}_R(p^2)$ goes toward $\Omega_R(p^2)$ in the supersymmetric limit of $x\to 0$.

There are also contributions of the gauge and gaugino fields to the gaugino propagator.
Due to supersymmetry, these contributions are the same as those of gauge and gaugino fields to the gauge
field propagator,
\be
\Sigma^{[2]}_R(p^2)=\Pi^{[2]}_R(p^2),
\ee
and then the full 2-point function for the massless gaugino field is given by
\be\label{eq:gauginor}
\Sigma_R(p^2)=\Sigma^{[1]}_R(p^2)+\Sigma^{[2]}_R(p^2).
\ee
\DOUBLEFIGURE {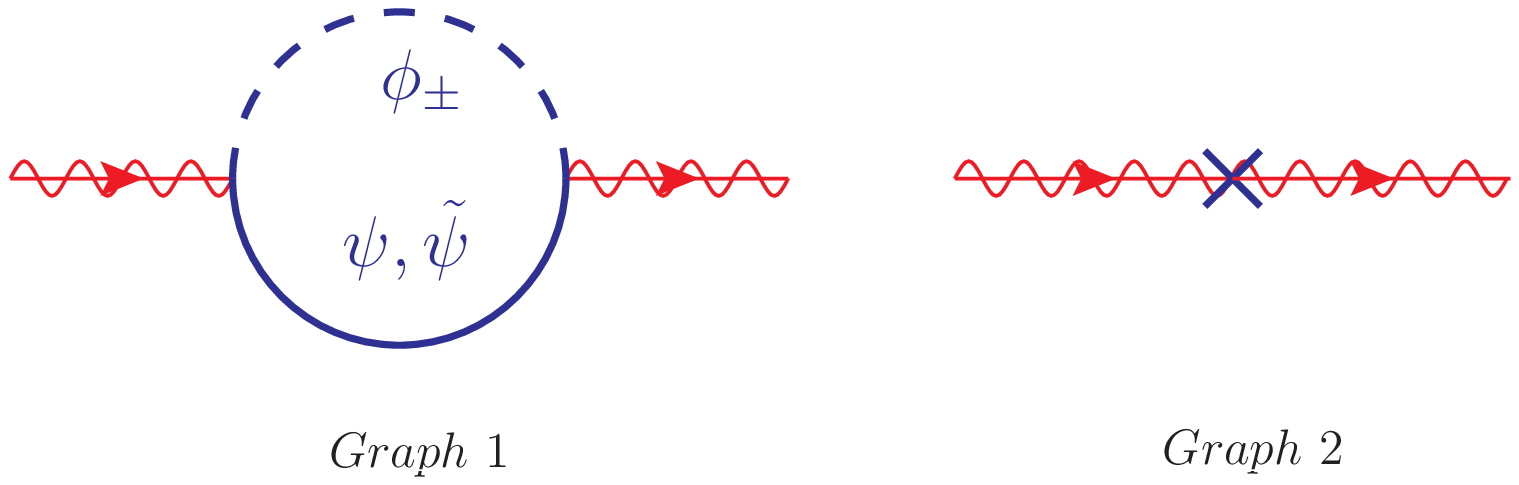,width=7 cm} {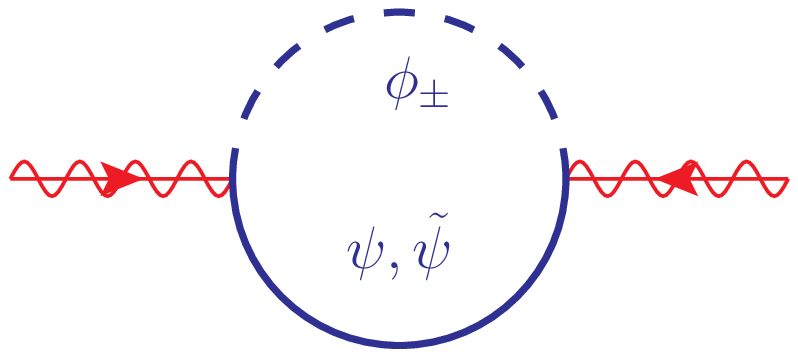,width=4 cm}
{The one-loop and counterterm corrections to the helicity-preserving gaugino propagator.
The blue plain and dotted lines in Graph 1 denote the messenger fermions and scalars, respectively,
while the cross in Graph 2  the counterterm vertex.\label{fig:sigma1}}
{The one-loop contribution to the helicity-flipping gaugino propagator.
The blue plain and dotted lines denote the messenger fermions and scalars, respectively.\label{fig:sigma2}}

By summing a geometric series of a massless propagator and a invariant function $\Sigma_R$
as depicted in fig.~\ref{fig:pi2} and fig.~\ref{fig:sigma1},
the full propagator for the massless gaugino field, as graphically shown in fig.~\ref{fig:sigma3},
is given by
\be
\frac{-i\bar\sigma^{\mu\dot\alpha\alpha}p_\mu}{p^2}\frac{1}{1-\Sigma_R(p^2)}
= \frac{-i\bar\sigma^{\mu\dot\alpha\alpha}p_\mu}{p^2}\big(1+\Sigma_R+\Sigma_R^2+\cdots\big).
\ee
\FIGURE {\epsfig{file=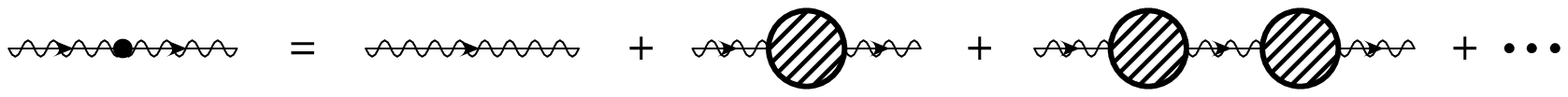, width=15 cm}
\caption{The full massless gaugino propagator in the basis of 2-component spinors.
The line with the dot on the left-hand side denotes the full propagator for a massless gaugino field,
and the single-shaded blobs the helicity-conserving 1PI self-energies as shown in fig.~\ref{fig:sigma1}.}\label{fig:sigma3}}

On the other hand, the graph in fig.~\ref{fig:sigma2} depicts the helicity changing propagator at one loop.
Ultraviolet divergences of the one-loop diagrams with different internal loop fields in fig.~\ref{fig:sigma2} cancel by itself
so that no counterterm is required. The second spinor 2-point function for the graph in fig.~\ref{fig:sigma2}
is at 1-loop level given by  $\epsilon_{\alpha\beta}i\mathcal{M}^{(1)}(p^2)$, where
\bea\label{eq:pgmass}
i{\cal M}^{(1)}(p^2) &=& -2M_0\frac{g^2 \mathbf{C}(\mathrm{R})}{16 \pi^2} \mu^{2\varepsilon}
\int \frac{d^d q}{(2\pi)^d}\bigg(\frac{1}{q^2-M^2_-}-\frac{1}{q^2-M^2_+}\bigg) \frac{1}{(p+q)^2-M^2_0}
\nonumber\\
&=& -2iM_0\frac{g^2 \mathbf{C}(\mathrm{R})}{16 \pi^2}
\int^1_0 dy \log\bigg[\frac{M^2_0+(1-y)F-y(1-y)p^2}{M^2_0-(1-y)F-y(1-y)p^2}\bigg]\nonumber\\
&=&-2iM_0\frac{g^2 \mathbf{C}(\mathrm{R})}{16 \pi^2}
\bigg\{\frac{x}{2\hat p^2_0}\log(1-x^2)-\frac{1}{2}\log\frac{1-x}{1+x}\nonumber\\
&&+\bigg(\frac{x}{\hat p^2_0}-1\bigg)\bigg[h\bigg(\frac{x-\hat p^2_0}{\sqrt{(x+\hat p^2_0)^2-4\hat p^2_0}}\bigg)
-h\bigg(\frac{x-\hat p^2_0}{\sqrt{(x-\hat p^2_0)^2-4\hat p^2_0}}\bigg)\bigg]\nonumber\\
&&+\bigg(\frac{x}{\hat p^2_0}+1\bigg)\bigg[h\bigg(\frac{x+\hat p^2_0}{\sqrt{(x-\hat p^2_0)^2-4\hat p^2_0}}\bigg)
-h\bigg(\frac{x+\hat p^2_0}{\sqrt{(x+\hat p^2_0)^2-4\hat p^2_0}}\bigg)\bigg]\bigg\}.
\eea
We note that ${\cal M}^{(1)}$ is independent of the renormalization scale $\mu$
and vanishing in the supersymmetric limit of $x\to 0$. For $p^2=0$,
it is reduced to the well-known gaugino mass parameter as in ref.~\cite{Martin:1996zb}.
\FIGURE {\epsfig{file=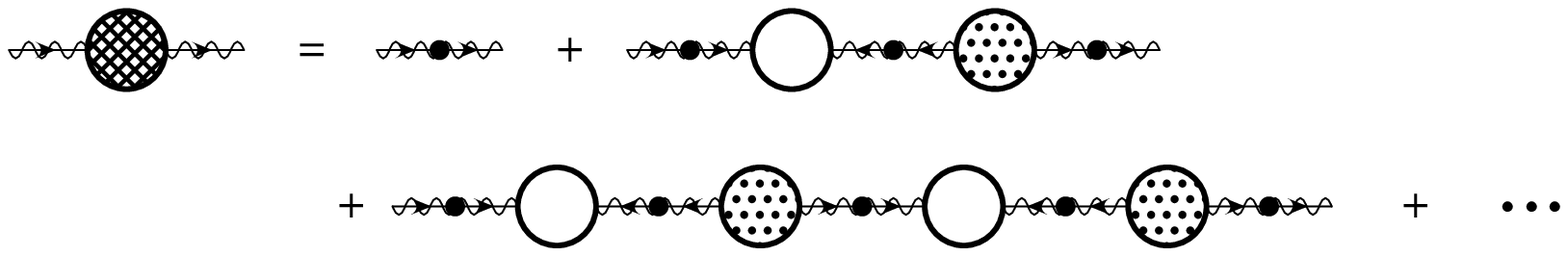, width=15.5 cm}
\caption{The full helicity-conserving propagator for the massive gaugino field.
The line with double-shaded blob on the left hand side denotes the full massive gaugino propagator,
and the line with black dots the full massless gaugino propagator as shown in fig.~\ref{fig:sigma3}.
The blank blobs denote the amplitude for the 1PI graphs as shown in fig.~\ref{fig:sigma2},
and the dot-shaded blobs its complex conjugate.}\label{fig:sigma4}}

We now consider two propagators for a massive gaugino field.
The {\it full} helicity-preserving propagator, as graphically shown in fig.~\ref{fig:sigma4}, is given by
\be
\frac{-i\bar\sigma^{\mu\dot\alpha\alpha}p_\mu}{p^2
-\frac{|\mathcal{M}|^2}{(1-\Sigma_R)^2}} \frac{1}{1-\Sigma_R}
=\frac{-i\bar\sigma^{\mu\dot\alpha\alpha}p_\mu}{p^2(1-\Sigma_R)}
\bigg\{1+\frac{|\mathcal{M}|^2}{p^2(1-\Sigma_R)^2}
+\bigg[\frac{|\mathcal{M}|^2}{p^2(1-\Sigma_R)^2}\bigg]^2
+\cdots\bigg\},
\ee
\FIGURE {\epsfig{file=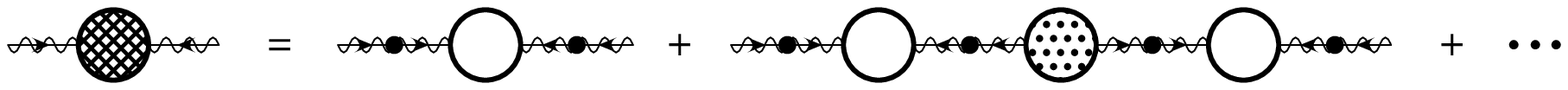, width=15.5 cm}
\caption{The full helicity-flipping propagator of a massive gaugino field.
The line with double-shaded blob denotes the full massive gaugino propagator,
and the line with black dots the full massless gaugino propagator as shown in fig.~\ref{fig:sigma3}.
The blank blobs denote the amplitude for the 1PI diagrams as shown in fig~\ref{fig:sigma2},
and the dot-shaded blobs its complex conjugate.}\label{fig:sigma5}}
while the {\it full} helicity-flipping propagator, as shown in fig.~\ref{fig:sigma5}, is given by
\be
\frac{i\frac{\mathcal{M}}{1-\Sigma_R}}{p^2-\frac{|\mathcal{M}|^2}{(1-\Sigma_R)^2}} \frac{1}{1-\Sigma_R}
=\frac{i\mathcal{M}}{p^2(1-\Sigma_R)^2}
\bigg\{1+\frac{|\mathcal{M}|^2}{p^2(1-\Sigma_R)^2}
+\bigg[\frac{|\mathcal{M}|^2}{p^2(1-\Sigma_R)^2}\bigg]^2
+\cdots\bigg\}.
\ee
We note that the wavefunction renormalization constants for both of the spinor propagators are the same as expected.
\subsection{Supersymmetric current correlators}
In the previous section we showed that $\Omega_R(p^2), \Pi^{[1]}_R(p^2)$ and $\Sigma^{[1]}_R(p^2)$ are,
once supersymmetry is spontaneously broken ($F\neq0$),
slightly different while ${\cal M}(p^2)$ is non-vanishing.
It is noteworthy that the expressions for them are valid
both for $p^2<F$ and for $\mu^2<F$, where supersymmetry is spontaneously broken.
For $p^2>F$, supersymmetry is restored
so that they are all equal and given by~\footnote{It is obtained from
eq.~(\ref{eq:omegar}) by putting $x=0$.}
\be
\Omega_R(p^2)= \Pi^{[1]}_R(p^2)= \Sigma^{[1]}_R(p^2)=
\frac{2g^2 \mathbf{C}(\mathrm{R})}{16 \pi^2}\bigg[\log\frac{\mu^2}{M^2_0}
+2-2h\bigg(\sqrt{\frac{\hat p^2_0}{4-\hat p^2_0}}\bigg)\bigg],
\ee
while ${\cal M}(p^2)$=0.
On the other hand, the expression for $\Pi^{[2]}(p^2)$ and $\Sigma^{[2]}(p^2)$
is valid {\it not only} for $p^2<F$ {\it but also} for $p^2>F$.
  
There are also higher order correlation functions involved with the vector superfield
and the renormalization of them would have to be taken into account for the complete analysis.
We will perform the computations of them in the future work.

We make a comment on the scale dependence of the correlation functions.
Whether or not supersymmetry is spontaneously broken in the messenger sector
the coefficients of the $\log\mu$ term in the correlation functions are the same.
The differences between supersymmetric and non-supersymmetric correlation functions
are only from finite terms.
\section{Gaugino masses}
We compute the {\it physical} or {\it renoramalized} gaugino mass,
which would be measured at the LHC.
As previously discussed, the 1PI graph, as shown in fig.~\ref{fig:sigma2}, gives a mass to the gaugino.
No ultraviolet divergences arise from the 1PI graph,
which is easily understood by comparing eq.~(\ref{eq:220}) with eq.~(\ref{eq:2212}).
For instance, ultraviolet divergences of one of the the graph, as shown in fig.~\ref{fig:sigma2}, is cancelled by another
with the minus sign, as in eq.~(\ref{eq:2212}).
As a result, the pole mass of gaugino is finite and given by
\be\label{eq:pgauginom}
\bar m_\lambda(p^2=\bar m^2_\lambda,\mu)=
\frac{\mathcal{M}(p^2=\bar m^2_\lambda)}{1-\Sigma_R(p^2=\bar m^2_\lambda,\mu)},
\ee
where ${\cal M}$ is the amplitude of the 1PI graph, and 
$(1-\Sigma_R)^{-1}$ is the gaugino wavefunction renormalization constant.
In order to get a physical mass we have to renormalize the gaugino wavefunction
so that we get rid of the $\mu$ dependence in eq.~(\ref{eq:pgauginom}).
But we do not specify the renormalization procedure here~\footnote{It is under current investigation~\cite{current}.}.

We can now expand eq.~(\ref{eq:pgauginom}) in a systematic way.
To the next-to-leading order (NLO) in $\alpha(=g^2/4\pi)$, the physical gaugino mass is given by,
\bea\label{eq:gauginomass}
\bar m_\lambda(p^2=\bar m_\lambda^2,\mu)&\simeq&\mathcal{M}^{(1)}(p^2=\bar m_\lambda^2)
+\mathcal{M}^{(2)}(p^2=\bar m_\lambda^2)\nonumber\\
&&+\mathcal{M}^{(1)}(p^2=\bar m_\lambda^2)\Sigma^{(1)}_R(p^2=\bar m_\lambda^2,\mu),
\eea
where ${\cal M}^{(1)}$ and $\Sigma^{(1)}_R$ are given by eq.~(\ref{eq:pgmass})
and~(\ref{eq:gauginor}), respectively, while $\mathcal{M}^{(2)}$ is the two-loop 1PI amplitude.
Recall that the number in the superscript parenthesis stands for the order of loops.
To perform the computation of complete NLO effects
we would have to consider {\it not only} the two-loop 1PI amplitude $\mathcal{M}^{(2)}$
{\it but also} the one-loop gaugino wavefunction renormalization.
 
To complete the renormalization procedures
we would have to take account of the gauge coupling renormalization as well.
We sketch how to do it but the details are under current investigation~\cite{current}.
The gauge coupling renormalization $Z_g$ is given as
\be
g_0=Z_gg\mu^\varepsilon,
\ee
where $g_0(g)$ is the bare(renormalized) coupling constant.
The straightforward computation of $Z_g$ is through the renormalization
of the gaugino-gaugino-gauge field vertex $Z_3$ such that
\be
Z_g=\frac{Z_3}{Z_v^{1/2} Z_\lambda},
\ee
where $Z_v(Z_\lambda)$ is the gauge(gaugino) field wavefunction renormalization.
The one-loop corrections to the gaugino-gaugino-gauge field vertex is shown in fig.~\ref{fig:vertex}. 
There are two kinds of corrections to the vertex: 
one comes from the gaugino and gauge field loops as in the graph 1 and 2 of  fig.~\ref{fig:vertex}
while the other from the messenger field loops as in the graph 3 and 4 of fig.~\ref{fig:vertex}.
\FIGURE {\epsfig{file=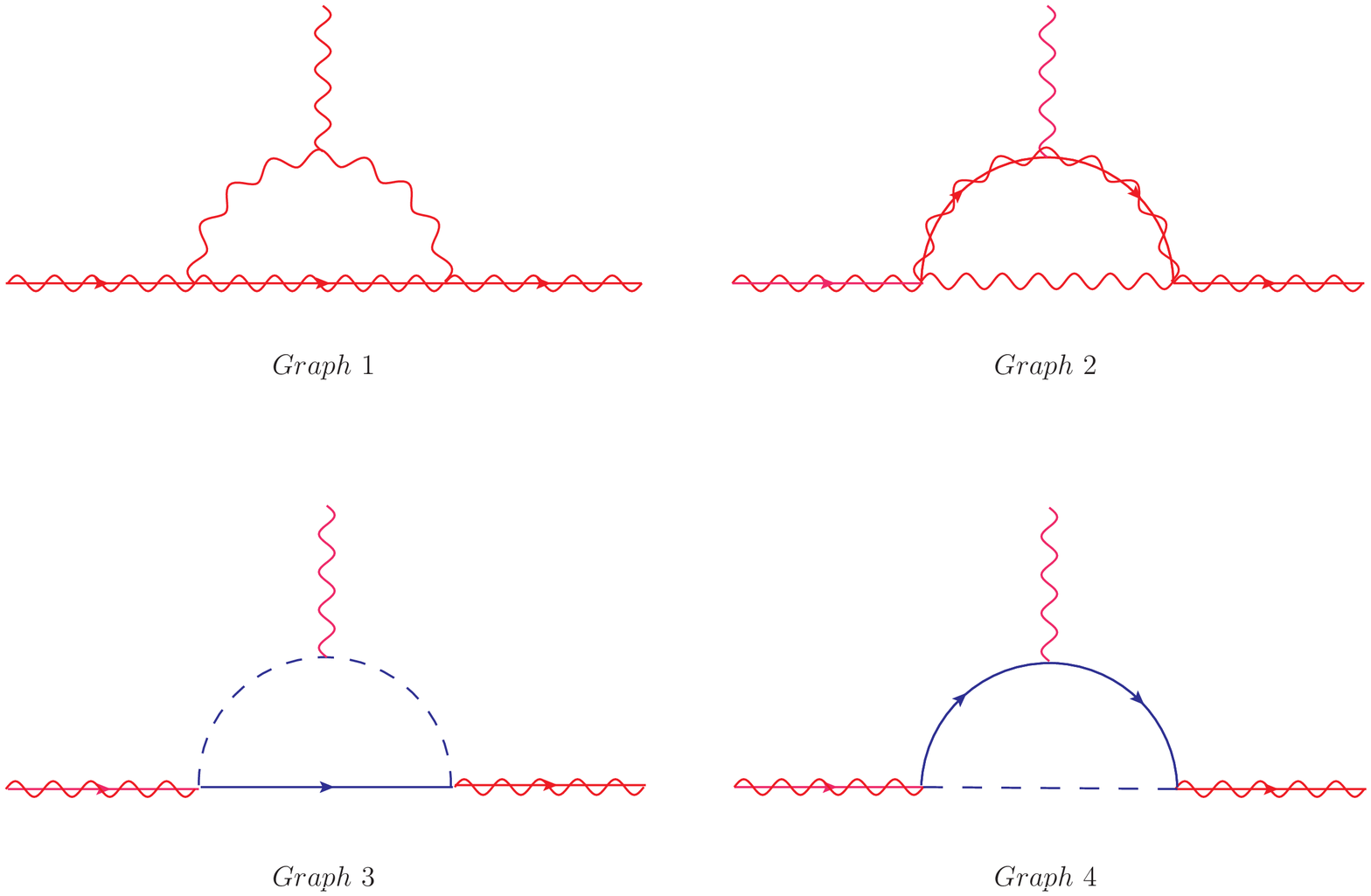, width=10 cm}
\caption{The one-loop corrections to the gaugino-gaugino-gauge field vertex.}\label{fig:vertex}}
\section{Sfermion masses}
We turn to the sfermion mass.
It is protected at one loop from quadratic divergences from gauge interactions
so the mass corrections begin to arise at two loop.
In fact, the previous discussion about renormalization of the gaugino mass is also applied to the sfermion mass.
The only difference is that the gaugino mass begins to arise at one loop while the sfermion
mass at two loop. Thus the NLO effects for the sfermion mass arise at three loop.

The full propagator of a sfermion is, graphically shown in fig.~\ref{fig:sfer1}, given
by a summing of a geometric series of a massless propagator and an amplitude
$-i\Xi$ for the 1PI graphs as,
\FIGURE {\epsfig{file=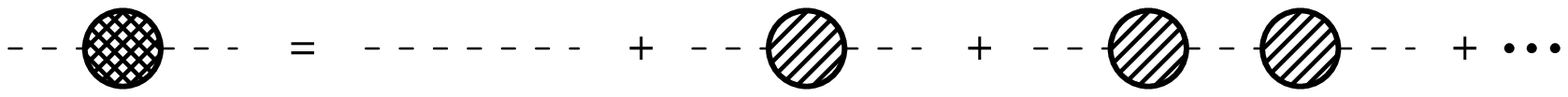, width=15.4 cm}
\caption{The full propagator for a massive sfermion.
The double-shaded blob represents its mass while the single-shaded blobs
are 1PI graphs shown in fig.~\ref{fig:sfer}.}\label{fig:sfer1}}
\be
\frac{i}{q^2-\Xi(q^2)}=\frac{i}{q^2}+\frac{i}{q^2}\frac{\Xi(q^2)}{i}\frac{i}{q^2}
+\frac{i}{q^2}\frac{\Xi(q^2)}{i}\frac{i}{q^2}\frac{\Xi(q^2)}{i}\frac{i}{q^2}
+\cdots,
\ee
where $q^\mu$ is the four momentum of the sfermion,
and $\Xi(q^2)$ is decomposed into two parts,
\be
\Xi(q^2,\mu)= q^2\xi(q^2,\mu)+\chi(\mu),
\ee
where $\chi$ is independent of $q^2$ but still depends on supersymmetry breaking parameters.
As a result, the resummation of the sfermion propagators can be decomposed into
the renormalized propagator and wavefunction renormalization constant,
\be
\frac{i}{q^2-\Xi}=\frac{i}{q^2-\frac{\chi}{1 - \xi}}\frac{1}{1 - \xi},
\ee
where the sfermion wavefunction renormalization constant is given as $(1 - \xi)^{-1}$,
and the pole mass of sfermion is
\be
\bar m^2_{\tilde f}(q^2,\mu)= \frac{\chi(\mu)}{1 - \xi(q^2,\mu)}.
\ee
To the NLO in $\alpha$, the physical sfermion mass is given by
\be
\bar m^2_{\tilde f}(q^2=\bar m^2_{\tilde f},\mu)\simeq \chi^{(2)}(\mu)
+\chi^{(2)}(\mu)\xi^{(1)}(q^2=\bar m^2_{\tilde f},\mu)+\chi^{(3)}(\mu),
\ee
where $\chi^{(2)}$ is the well-known sfermion mass parameter as in ref.~\cite{Martin:1996zb}.
Of course, we would have to take account of renormalization as well but we will not do it in the letter.

In the context of GGM, the 1PI amplitude is given by the graphs as shown in fig.~\ref{fig:sfer},
\FIGURE {\epsfig{file=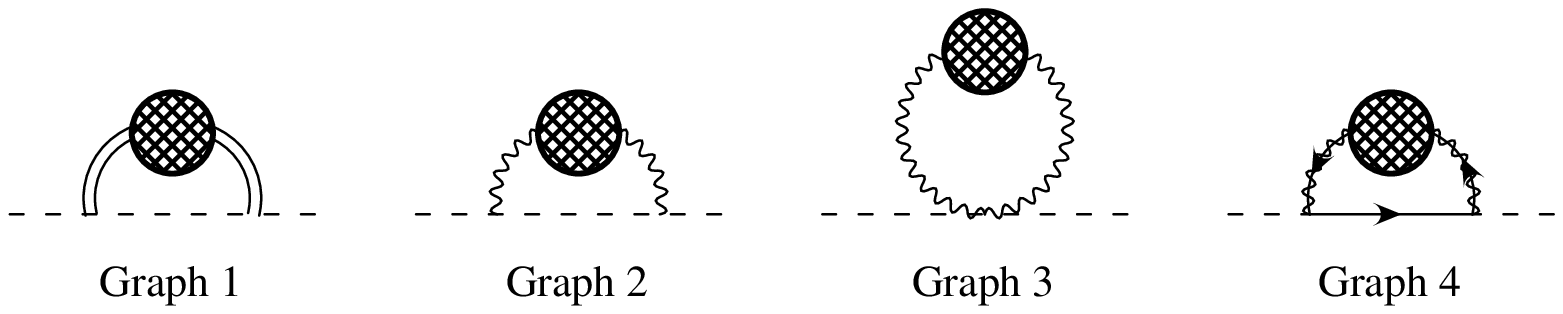, width=15 cm}
\caption{The 1PI graphs for a sfermion mass.
The lines with double-shaded blobs stand for the full propagators for the components
of the vector superfield.}\label{fig:sfer}}
\be -i\Xi(q^2)= g^2\mathbf{C}(\mathrm{R}) \sum_{j=1}^4{\cal A}_j, \ee
where
\bea
{\cal A}_1(q^2)&=& \int^F \frac{\mathrm{d}^4p}{(2\pi)^4} \frac{1}{(p+q)^2}\frac{1}{1-\Omega_R(p^2)},\\
{\cal A}_2(q^2)&=& -4 \int ^F\frac{\mathrm{d}^4p}{(2\pi)^4}\frac{q^2p^2-(p\cdot q)^2}{(p^2)^2(p+q)^2}
\frac{1}{1-\Pi_R(p^2)},\\
{\cal A}_3&=& 3 \int^F \frac{\mathrm{d}^4p}{(2\pi)^4} \frac{1}{p^2}\frac{1}{1-\Pi_R(p^2)},\\
{\cal A}_4(q^2)&=& -4\int^F \frac{\mathrm{d}^4p}{(2\pi)^4} \frac{p^2+p\cdot q}{(p+q)^2}
\frac{1}{p^2-\frac{|\mathcal{M}(p^2)|^2}{(1-\Sigma_R)^2}}\frac{1}{1-\Sigma_R(p^2)}.
\eea
Note that only ${\cal A}_3$ is independent of $q^2$ so that it does not contribute
to the wavefunction renormalization constant of the sfermion.
On the other hand, ${\cal A}_{1,2,4}$ contribute not only to the wavefunction renormalization constant
of the sfermion but also to the mass of the sfermion.

Before closing this section, we would like to make a comment.
As mentioned in the previous section, 3-point functions or higher order correlation functions
of a vector superfield also lead to radiative corrections to the sfermion mass.
These contributions can be systematically taken into account in parallel with a renormalization procedure,
and will produce a result more approximate to the physical mass measured at the detectors of the LHC.
Moreover, once parity conservation under $V\to -V$ is introduced, odd order correlation functions vanish
so the next leading order correlation function begins at the fourth.     
We will pursue computation of higher order correlation functions in the future~\cite{current}.

\section{Summary and outlook}
We revisited the GGM, in particular by taking the minimal gauge mediation as a specific example.
The current superfield of an ungauged Wess-Zumino model is derived from
an infinitesimal global transformation of the model.
The 2-point functions of the current superfield is computed using (anti-)commutation relations
among the component fields. The Fourier transformations of the scalar, first spinor and vector 2-point functions
yield the vacuum polarization tensor and its two supersymmetric counterparts.
Their invariant functions correspond to the wavefunction renormalization constants of
the component fields of a vector superfield so we understand why these invariant functions
are all the same in supersymmetric theories.
For spontaneously broken supersymmetric theories, these invariant functions
are all slightly different, and so do the wavefunction renormalization constants of
the component fields of a vector superfield.
We computed these functions at one loop using dimensional regularization.

We have shown that the physical mass of a gaugino is expressed in terms of the 1PI graphs
and the wavefunction renormalization constants of the gaugino fields.
The physical masses of squarks and sleptons are also expressed in terms of the wavefunction
renormalization constants of the gaugino fields and the gaugino masses.
Higher order correlation functions should be taken into account in order to make the effects of higher
order loop of the 1PI graphs included.

The linear sigma model as an inceptive model of the GGM can be generalized to the nonlinear sigma model.
In superspace notation, it implies a non-minimal K\"{a}hler potential.
One can regard a non-minimal K\"{a}hler potential as an effective K\"{a}hler potential
resulting from integrating out heavy superfields.
In this regards, the GGM may help merge the gauge mediation with other mediation mechanisms.

\section{Acknowledgements}
We would like to thank Kwang Sik Jeong, Sung Jay Lee and Ann Nelson for useful comments and discussions.
This research was supported by Basic Science Research Program
through the National Research Foundation of Korea(NRF) funded
by the Ministry of Education, Science and Technology, under contract 2010-0012779.
\appendix

\section{Calculation of 2-point functions}\label{app1}
In the Appendix we express four 2-point functions in terms of the current superfield as in (\ref{eq:expcur2}).
To begin with, we write down the free field 2-point functions, the propagators,
which are necessary to express the 2-point functions of the current superfield:
\bea
\langle 0|\mbox{T} [\phi_\pm(x)\phi^\ast_\pm(0)]|0\rangle &=& i\Delta_F(x;M_\pm),\\
\langle 0|\mbox{T}[\psi_\alpha(x) \psi^\beta(0)]|0\rangle &=&\langle 0|\mbox{T}[\tilde\psi_\alpha(x) \tilde\psi^\beta(0)]|0\rangle
=i\delta^{\,\,\,\beta}_\alpha M_0 \Delta_F(x;M_0),\label{eq:fpf11}\\
\langle 0|\mbox{T}[\bar\psi^{\dot\alpha}(x) \bar\psi_{\dot\beta}(0)]|0\rangle
&=&\langle 0|\mbox{T}[\bar{\tilde\psi}^{\dot\alpha}(x) \bar{\tilde\psi}_{\dot\beta}(0)]|0\rangle
=i\delta^{\dot\alpha}_{\,\,\dot\beta} M_0\Delta_F(x;M_0),\label{eq:fpf12}\\
\langle 0|\mbox{T}[\psi_\alpha(x) \bar\psi_{\dot\beta}(0)]|0\rangle 
&=&\langle 0|\mbox{T}[\tilde\psi_\alpha(x) \bar{\tilde\psi}_{\dot\beta}(0)]|0\rangle 
=\sigma^\mu_{\alpha\dot\beta} \partial_\mu \Delta_F(x;M_0),\label{eq:fhf22}
\eea
where $\Delta_F(x;M)$ is the propagator for a scalar field with mass $M$,
\be
\Delta_F(x;M) \equiv\frac{1}{\Box-M^2}.
\ee
Note that there are two kinds of fermion propagators with respect to helicity:
one preserves it as in eq.~(\ref{eq:fhf22})
while the other flips it as in eq.~(\ref{eq:fpf11}) and (\ref{eq:fpf12}). 

We first evaluate the scalar 2-point function of the current superfield:
\be\label{eq:scalar2}\begin{aligned}
\langle J^a(x)J^a(0)\rangle 
&\equiv \langle 0|\mbox{T}\big\{ J^a(x)J^a(0)\big\}|0\rangle\\
&= \langle0| \mbox{T}\big\{ \big(\phi^\dagger_+ T^a\phi_-(x)+\phi^\dagger_-T^a\phi_+(x)\big)
\big(\phi^\dagger_+ T^a\phi_-(0)+\phi^\dagger_-T^a\phi_+(0)\big)\big\}|0 \rangle\\
&= \langle 0|\mbox{T}\big\{\phi^\dagger_+(x)T^a\phi_-(x)\,\phi^\dagger_-(0) T^a \phi_+(0)\big\}|0\rangle
+\langle 0|\mbox{T}\big\{\phi^\dagger_-(x)T^a\phi_+(x)\,\phi^\dagger_+(0) T^a \phi_-(0)\big\}|0\rangle\\
&= 2\langle0| \mbox{T}\big\{\phi^{\ast b}_+(x)\phi_{+e}(0)\big\}|0\rangle 
\langle0| \mbox{T}\big\{\phi^\ast_{-c}(x)\phi^{\,\,d}_-(0)\big\}|0\rangle T^{\boldsymbol{a}\,c}_{\,\,b}T^{\boldsymbol{a}\,e}_{\,\,d}\\
&= 2\delta^{\,\,b}_e i\Delta_F(x;M_+) \delta^{\,\,d}_c i\Delta_F(x;M_-) T^{\boldsymbol{a}\,c}_{\,\,b}T^{\boldsymbol{a}\,e}_{\,\,d}\\
&= -2 \mathbf{C}(\mathrm{R}) \Delta_F(x;M_+) \Delta_F(x;M_-),
\end{aligned}\ee
where $\mathbf{C}(\mathrm{R})$ is the quadratic invariant of the fundamental representation,
\be\mbox{Tr} [T^aT^b] =   \mathbf{C}(\mathrm{R})\delta^{ab}.\ee
Note that (\ref{eq:scalar2}) is symmetric under the exchange $M_+\leftrightarrow M_-$.

Next we turn to the first spinor 2-point function
\be\label{eq:1spinor2}\begin{aligned}
\langle j^a_\alpha(x)\bar j^a_{\dot\alpha}(0)\rangle
&\equiv \langle 0|\mbox{T}\big\{ j^a_\alpha(x)\bar j^a_{\dot\alpha}(0)\big\}|0\rangle\\
&=\langle0| \mbox{T}\big\{[\phi^\dagger_+(x)+\phi^\dagger_-(x)] T^a\psi_\alpha(x)\,
\bar\psi_{\dot\alpha}(0) T^a [\phi_+(0)+\phi_-(0)]\big\}|0\rangle\\
&\quad+\langle0|\mbox{T}\big\{\tilde\psi_\alpha (x)T^a [\phi_+(x)-\phi_-(x)]\,
[\phi^\dagger_+(0)-\phi^\dagger_-(0)]T^a\bar{\tilde\psi}_{\dot\alpha}(0)\big\}|0\rangle\\
&=  T^{\boldsymbol{a}\,c}_{\,\,b}T^{\boldsymbol{a}\,e}_{\,\,d}\bigg\{
\big[\langle 0|\mbox{T}\big\{\phi^{\ast b}_+(x)\phi_{+e}(0)\big\}|0\rangle 
\langle 0|\mbox{T}\big\{\psi_{\alpha c}(x)\bar\psi^d_{\dot\alpha}(0)\big\}|0\rangle
+(\phi_+\to \phi_-)\big]\\
&\qquad\qquad+\langle 0|\mbox{T}\big\{\tilde\psi^b_\alpha(x)\bar{\tilde\psi}_{\dot\alpha e}(0)\big\}|0\rangle
\langle0|\mbox{T}\big\{\phi_{+c}(x)\phi^{\ast d}_+(0)\big\}|0\rangle +(\phi_+\to \phi_-)\big]\bigg\}\\
&= 2\mathbf{C}(\mathrm{R})\big[i\Delta_F(x;M_+)+i\Delta_F(x;M_-)\big]
\sigma_{\alpha\dot\alpha}^\mu \partial_\mu\Delta_F(x;M_0).
\end{aligned}\ee
Note that the first spinor 2-point function is a sum of product of one scalar and one helicity-preserving fermion propagators.
Moreover, it is symmetric under the exchange $M_+\leftrightarrow M_-$.

Next we turn to the evaluation of the second spinor 2-point function
\be\label{eq:2spinor2}\begin{aligned}
\langle j^a_\alpha(x) j^a_\beta(0)\rangle
&\equiv \langle 0|\mbox{T}\big\{j^a_\alpha(x) j^a_\beta(0)\big\}|0\rangle\\
&= \langle 0|\mbox{T}\big\{[\phi^\dagger_+(x)+\phi^\dagger_-(x)]T^a\psi_\alpha(x)\,
\tilde\psi_\beta(0)T^a[\phi_+(0)-\phi_-(0)]\big\}|0\rangle\\
&\quad+\langle 0|\mbox{T}\big\{[\tilde\psi_\alpha(x)T^a[\phi_+(x)-\phi_-(x)]\,
[\phi^\dagger_+(0)+\phi^\dagger_-(0)]T^a\psi_\beta(x)\big\}|0\rangle\\
&=T^{\boldsymbol{a}\,c}_{\,\,b}T^{\boldsymbol{a}\,e}_{\,\,d}\bigg\{
\big[\langle0|\mbox{T}\big\{\phi^{\ast b}_+(x)\phi_{+e}(0)]\big\}|0\rangle
-(\phi_+\to\phi_-)\big\}|0\rangle\big]
\langle 0|\mbox{T}\big\{\psi_{\alpha c}(x)\tilde\psi^d_\beta(0)\big\}|0\rangle\big]\\
&\qquad\qquad\quad+
\big[\langle0|\mbox{T}\big\{\phi_{+c}(x)\phi^{\ast d}_+(0)\big\}|0\rangle
-(\phi_+\to\phi_-)\big]
\langle 0|\mbox{T}\big\{\tilde\psi^b_{\alpha}(x)\psi_{\beta e}(0)\big\}|0\rangle\big]\bigg\}\\
&=2 \mathbf{C}(\mathrm{R})[\Delta_F(x;M_+)-\Delta_F(x;M_-)]\varepsilon_{\alpha\beta}M_0\Delta_F(x;M_0).
\end{aligned}\ee
Note that (\ref{eq:2spinor2}) is antisymmetric not only under the exchange $M_+\leftrightarrow M_-$
but also under the exchange $\alpha\leftrightarrow\beta$.

Next we compute the bosonic vector 2-point function
\be\label{eq:bvector2}\begin{aligned}
\langle j^{aB}_\mu (x) j^{aB}_\nu (0)\rangle 
&\equiv \langle 0|\mbox{T}\big\{ j^{aB}_\mu (x) j^{aB}_\nu (0)\big\}|0\rangle \\
&=- \langle 0|\mbox{T}\big\{\big(\phi^\dagger_+T^a\overleftrightarrow{\partial_\mu} \phi_-+\phi^\dagger_- T^a\overleftrightarrow{\partial_\mu} \phi_+ \big)(x) \big(\phi^\dagger_+ T^a\overleftrightarrow{\partial_\nu} \phi_- 
+\phi^\dagger_- T^a\overleftrightarrow{\partial_\nu} \phi_+ \big)(0)\big\}|0\rangle\\
&= -\langle 0|\mbox{T}\big\{\big(\phi^\dagger_+ T^a\overleftrightarrow{\partial_\mu} \phi_- \big)(x)
\big(\phi^\dagger_- T^a\overleftrightarrow{\partial_\nu} \phi_+\big)(0)\big\}|0\rangle
+(\phi_+\leftrightarrow \phi_-)\\
&= 2T^{\boldsymbol{a}\,c}_{\,\,b}T^{\boldsymbol{a}\,e}_{\,\,d}
\big(\langle 0|\big\{\partial_\mu \phi^{\ast b}_+(x) \partial_\nu \phi_{+e}(0)\big\}|0\rangle
\langle 0|\mbox{T}\big\{\phi_{-c}(x) \phi^{\ast d}_-(0)\big\}|0\rangle\\
&\qquad\qquad\!\!- \langle 0|\mbox{T}\big\{\partial_\mu \phi^{\ast b}_+ (x) \phi_{+e}(0)\big\}|0\rangle
\langle0|\mbox{T}\big\{\phi_{-c}(x) \partial_\nu\phi^{\ast d}_-(0)\big\}|0\rangle\\
&\qquad\qquad\!\!+ \langle0|\mbox{T}\big\{ \phi^{\ast b}_+(x) \phi_{+e}(0)\big\}|0\rangle
\langle0|\mbox{T}\big\{\partial_\mu\phi_{-c}(x) \partial_\nu\phi^{\ast d}_-(0)\big\}|0\rangle\\
&\qquad\qquad\!\!- \langle0|\mbox{T}\big\{ \phi^{\ast b}_+ (x) \partial_\nu\phi_{+e}(0)\big\}|0\rangle
\langle0|\mbox{T}\big\{\partial_\mu\phi_{-c}(x)\phi^{\ast d}_-(0)\big\}|0\rangle\big)\\
&=2 \mathbf{C}(\mathrm{R})\big[\Delta_F(x;M_-)\partial_\mu\partial_\nu \Delta_F(x;M_+)
-\partial_\mu\Delta_F(x;M_+)\partial_\nu\Delta_F(x;M_-)\\
&\qquad\qquad+\Delta_F(x;M_+)\partial_\mu\partial_\nu \Delta_F(x;M_-)
- \partial_\mu\Delta_F(x;M_-)\partial_\nu\Delta_F(x;M_+)\big].
\end{aligned}\ee
Note that (\ref{eq:bvector2}) is symmetric not only under the exchange $M_+\leftrightarrow M_-$
but also under the exchange $\mu\leftrightarrow\nu$.

We also compute the fermionic vector 2-point function
\be
\label{eq:fvector2}
\begin{aligned}
\langle j^{aF}_\mu (x) j^{aF}_\nu (0)\rangle 
&\equiv \langle 0|\mbox{T}\big\{ j^{aF}_\mu (x) j^{aF}_\nu (0)\big\}|0\rangle \\
&= \langle 0|\mbox{T}\big\{\big[\bar\psi(x) \bar\sigma_\mu T^a\psi(x)+\tilde\psi(x)\sigma_\mu T^a\bar{\tilde\psi}(x)\big]
\big[\bar\psi(0) \bar\sigma_\nu T^a\psi(0)+\tilde\psi(0)\sigma_\nu T^a\bar{\tilde\psi}(0)\big]\big\}|0\rangle\\
&=T^{\boldsymbol{a}\,c}_{\,\,b}T^{\boldsymbol{a}\,e}_{\,\,d}\bar\sigma_\mu^{\dot\alpha\alpha}\bar\sigma_\nu^{\dot\beta\beta}
\langle 0|\mbox{T}\big\{\bar \psi^b_{\dot\alpha}(x)\psi_{\beta e}(0)\big\}|0\rangle
\langle 0|\mbox{T}\big\{\psi_{\alpha c}(x)\bar\psi^d_{\dot\beta}(0)\big\}|0\rangle\\
&\,\,+T^{\boldsymbol{a}\,c}_{\,\,b}T^{\boldsymbol{a}\,e}_{\,\,d}\bar\sigma_\mu^{\dot\alpha\alpha}\sigma_{\nu\beta\dot\beta}
\langle 0|\mbox{T}\big\{\bar \psi^b_{\dot\alpha}(x)\bar{\tilde\psi}^{\dot\beta}_e(0)\big\}|0\rangle
\langle 0|\mbox{T}\big\{\psi_{\alpha c}(x)\tilde\psi^{\beta d}(0)\big\}|0\rangle\\
&\,\,+T^{\boldsymbol{a}\,c}_{\,\,b}T^{\boldsymbol{a}\,e}_{\,\,d}\sigma_{\mu\alpha\dot\alpha}\bar\sigma_\nu^{\dot\beta\beta}
\langle 0|\mbox{T}\big\{\tilde\psi^{\alpha b}(x)\psi_{\beta e}(0)\big\}|0\rangle
\langle 0|\mbox{T}\big\{\bar{\tilde\psi}_c^{\dot\alpha}(x)\bar\psi^d_{\dot\beta}(0)\big\}|0\rangle\\
&\,\, +T^{\boldsymbol{a}\,c}_{\,\,b}T^{\boldsymbol{a}\,e}_{\,\,d}\sigma_{\mu\alpha\dot\alpha}\sigma_{\nu\beta\dot\beta}
\langle 0|\mbox{T}\big\{\tilde\psi^{\alpha b}(x)\bar{\tilde\psi}_e^{\dot\beta}(0)\big\}|0\rangle
\langle 0|\mbox{T}\big\{\bar{\tilde\psi}_c^{\dot\alpha}(x)\tilde\psi^{\beta d}(0)\big\}|0\rangle\\
&=\mathbf{C}(\mathrm{R})\bar\sigma_\mu^{\dot\alpha\alpha}\bar\sigma_\nu^{\dot\beta\beta}
\sigma^\lambda_{\beta\dot\alpha}\partial_\lambda\Delta_F(x;M_0)
\sigma^\rho_{\alpha\dot\beta}\partial_\rho\Delta_F(x;M_0)\\
&\,\,+\mathbf{C}(\mathrm{R})\bar\sigma_\mu^{\dot\alpha\alpha}\sigma_{\nu\beta\dot\beta}
\delta^{\dot\beta}_{\dot\alpha} i M_0\Delta_F(x;M_0) \delta^\beta_\alpha iM_0\Delta_F(x;M_0)\\
&\,\,+\mathbf{C}(\mathrm{R})\sigma_{\mu\alpha\dot\alpha}\bar\sigma_\nu^{\dot\beta\beta}
\delta^\alpha_\beta i M_0\Delta_F(x;M_0) \delta^{\dot\alpha}_{\dot\beta}i M_0\Delta_F(x;M_0)\\
&\,\,+\mathbf{C}(\mathrm{R})\sigma_{\mu\alpha\dot\alpha}\sigma_{\nu\beta\dot\beta}
\bar\sigma^{\lambda\dot\beta\alpha}\partial_\lambda\Delta_F(x;M_0)
\bar\sigma^{\rho\dot\alpha\beta}\partial_\rho\Delta_F(x;M_0)\\
&=2 \mathbf{C}(\mathrm{R})\big[2\eta_{\mu\nu}\big(-\partial_\rho \Delta_F(x;M_0)\partial^\rho \Delta_F(x;M_0)
+M_0^2\Delta^2_F(x;M_0)\big) \\
&\qquad\qquad\,\,+ 4\partial_\mu \Delta_F(x;M_0)\partial_\nu \Delta_F(x;M_0) \big].
\end{aligned}\ee
Note that (\ref{eq:fvector2}) is symmetric under the exchange $\mu\leftrightarrow\nu$.
The term proportional to $M_0^2$ corresponds to a product of two helicity-flipping fermion propagators
while the other two terms do to a product of two helicity-conserving fermion propagators.

One looks up the Fourier transformation of these 2-point functions in ref.~\cite{Meade:2008wd,Marques:2009yu}.
\section{Loop corrections to sfermion mass}
In this Appendix we compute radiative corrections to the sfermion mass as shown in fig.~\ref{fig:sfer}.
\be
\begin{aligned}
(Graph~1)&=\int \frac{\mathrm{d}^4p}{(2\pi)^4} \frac{i}{(p+q)^2} (igT^a)\frac{i}{1-\Omega(p^2)}(igT^a)\\
&=g^2\mathbf{C}(\mathrm{R})\int \frac{\mathrm{d}^4p}{(2\pi)^4}\frac{1}{(p+q)^2}\frac{1}{1-\Omega(p^2)},\\
(Graph~2)&=\int \frac{\mathrm{d}^4p}{(2\pi)^4}[i(p+2q)_\mu gT^a]\frac{i}{(p+q)^2}
[i(p+2q)_\nu g T^a] \frac{-i}{p^2} \frac{\eta^{\mu\nu}-p^\mu p^\nu/p^2}{1-\Pi(p^2)}\\
&=-4g^2\mathbf{C}(\mathrm{R})\int \frac{\mathrm{d}^4p}{(2\pi)^4} 
\frac{p^2q^2-(p\cdot q)^2}{(p^2)^2(p+q)^2}\frac{1}{1-\Pi(p^2)},\\
(Graph~3)&=\int \frac{\mathrm{d}^4p}{(2\pi)^4}(i\eta_{\mu\nu}g^2 T^aT^a)
\frac{-i}{p^2}\frac{\eta^{\mu\nu}-p^\mu p^\nu/p^2}{1-\Pi(p^2)} \\
&=3g^2\mathbf{C}(\mathrm{R}) \int \frac{\mathrm{d}^4p}{(2\pi)^4}\frac{1}{p^2}\frac{1}{1-\Pi(p^2)},\\
(Graph~4)&=-\int \frac{\mathrm{d}^4p}{(2\pi)^4}
(\sqrt{2}gT^a)\frac{i\sigma^\mu_{\alpha\dot\alpha} (p+q)_\mu}{(p+q)^2}(-\sqrt{2}gT^a)
\frac{-i\bar\sigma^{\nu\dot\alpha\alpha}p_\nu}{p^2
-\frac{|\mathcal{M}|^2}{(1-\Sigma(p^2))^2}} \frac{1}{1-\Sigma(p^2)}\\
&=-4g^2 \mathbf{C}(\mathrm{R}) \int \frac{\mathrm{d}^4p}{(2\pi)^4} \frac{p^2+p\cdot q}{(p+q)^2}
\frac{1}{p^2-\frac{|\mathcal{M}|^2}{(1-\Sigma(p^2))^2}}\frac{1}{1-\Sigma (p^2)}.
\end{aligned}
\ee

\end{document}